\documentclass[12pt]{article}
\usepackage{amssymb}
\usepackage{graphicx}
\begin{document}
\newcommand{\beq}{\begin{equation}}
\newcommand{\eeq}{\end{equation}}
\newcommand{\beqa}{\begin{eqnarray}}
\newcommand{\eeqa}{\end{eqnarray}}
\newcommand{\beqar}{\begin{eqnarray*}}
\newcommand{\eeqar}{\end{eqnarray*}}
\newcommand{\al}{\alpha}
\newcommand{\be}{\beta}
\newcommand{\del}{\delta}
\newcommand{\D}{\Delta}
\newcommand{\eps}{\epsilon}
\newcommand{\ga}{\gamma}
\newcommand{\Ga}{\Gamma}
\newcommand{\ka}{\kappa}
\newcommand{\nn}{\nonumber}
\newcommand{\inn}{\!\cdot\!}
\newcommand{\h}{\eta}
\newcommand{\ii}{\iota}
\newcommand{\kk}{\varphi}
\newcommand\F{{}_3F_2}
\newcommand{\la}{\lambda}
\newcommand{\La}{\Lambda}
\newcommand{\na}{\prt}
\newcommand{\Om}{\Omega}
\newcommand{\om}{\omega}
\newcommand{\p}{\phi}
\newcommand{\sig}{\sigma}
\renewcommand{\t}{\theta}
\newcommand{\z}{\zeta}
\newcommand{\ssc}{\scriptscriptstyle}
\newcommand{\eg}{{\it e.g.,}\ }
\newcommand{\ie}{{\it i.e.,}\ }
\newcommand{\labell}[1]{\label{#1}} 
\newcommand{\reef}[1]{(\ref{#1})}
\newcommand\prt{\partial}
\newcommand\veps{\varepsilon}
\newcommand{\pol}{\varepsilon}
\newcommand\vp{\varphi}
\newcommand\ls{\ell_s}
\newcommand\cF{{\cal F}}
\newcommand\cA{{\cal A}}
\newcommand\cS{{\cal S}}
\newcommand\cT{{\cal T}}
\newcommand\cV{{\cal V}}
\newcommand\cL{{\cal L}}
\newcommand\cM{{\cal M}}
\newcommand\cN{{\cal N}}
\newcommand\cG{{\cal G}}
\newcommand\cH{{\cal H}}
\newcommand\cI{{\cal I}}
\newcommand\cJ{{\cal J}}
\newcommand\cl{{\iota}}
\newcommand\cP{{\cal P}}
\newcommand\cQ{{\cal Q}}
\newcommand\cg{{\it g}}
\newcommand\cR{{\cal R}}
\newcommand\cB{{\cal B}}
\newcommand\cO{{\cal O}}
\newcommand\tcO{{\tilde {{\cal O}}}}
\newcommand\bg{\bar{g}}
\newcommand\bb{\bar{b}}
\newcommand\bH{\bar{H}}
\newcommand\bF{\bar{F}}
\newcommand\bX{\bar{X}}
\newcommand\bK{\bar{K}}
\newcommand\bA{\bar{A}}
\newcommand\bZ{\bar{Z}}
\newcommand\bxi{\bar{\xi}}
\newcommand\bphi{\bar{\phi}}
\newcommand\bpsi{\bar{\psi}}
\newcommand\bprt{\bar{\prt}}
\newcommand\bet{\bar{\eta}}
\newcommand\btau{\bar{\tau}}
\newcommand\hF{\hat{F}}
\newcommand\hA{\hat{A}}
\newcommand\hT{\hat{T}}
\newcommand\htau{\hat{\tau}}
\newcommand\hD{\hat{D}}
\newcommand\hf{\hat{f}}
\newcommand\hg{\hat{g}}
\newcommand\hp{\hat{\phi}}
\newcommand\hi{\hat{i}}
\newcommand\ha{\hat{a}}
\newcommand\hb{\hat{b}}
\newcommand\hQ{\hat{Q}}
\newcommand\hP{\hat{\Phi}}
\newcommand\hS{\hat{S}}
\newcommand\hX{\hat{X}}
\newcommand\tL{\tilde{\cal L}}
\newcommand\hL{\hat{\cal L}}
\newcommand\tG{{\widetilde G}}
\newcommand\tg{{\widetilde g}}
\newcommand\tphi{{\widetilde \phi}}
\newcommand\tPhi{{\widetilde \Phi}}
\newcommand\te{{\tilde e}}
\newcommand\tk{{\tilde k}}
\newcommand\tf{{\tilde f}}
\newcommand\ta{{\tilde a}}
\newcommand\tb{{\tilde b}}
\newcommand\tR{{\tilde R}}
\newcommand\teta{{\tilde \eta}}
\newcommand\tF{{\widetilde F}}
\newcommand\tK{{\widetilde K}}
\newcommand\tE{{\widetilde E}}
\newcommand\tpsi{{\tilde \psi}}
\newcommand\tX{{\widetilde X}}
\newcommand\tD{{\widetilde D}}
\newcommand\tO{{\widetilde O}}
\newcommand\tS{{\tilde S}}
\newcommand\tB{{\widetilde B}}
\newcommand\tA{{\widetilde A}}
\newcommand\tT{{\widetilde T}}
\newcommand\tC{{\widetilde C}}
\newcommand\tV{{\widetilde V}}
\newcommand\thF{{\widetilde {\hat {F}}}}
\newcommand\Tr{{\rm Tr}}
\newcommand\tr{{\rm tr}}
\newcommand\STr{{\rm STr}}
\newcommand\hR{\hat{R}}
\newcommand\MZ{\mathbb{Z}}
\newcommand\MR{\mathbb{R}}
\newcommand\M[2]{M^{#1}{}_{#2}}

\newcommand\bS{\textbf{ S}}
\newcommand\bI{\textbf{ I}}
\newcommand\bJ{\textbf{ J}}

\begin{titlepage}
\begin{center}

\vskip 2 cm
{\LARGE \bf    Four-Derivative Yang-Mills Couplings in \\  \vskip 0.25 cm Heterotic Theory through T-Duality 
 }\\
\vskip 1.25 cm
 Mohammad R. Garousi\footnote{garousi@um.ac.ir}
 
\vskip 1 cm
{{\it Department of Physics, Faculty of Science, Ferdowsi University of Mashhad\\}{\it P.O. Box 1436, Mashhad, Iran}\\}
\vskip .1 cm
 \end{center}
\begin{abstract}

This study delves into the dimensional reduction of the classical effective action of heterotic string theory on a circle, along with its T-duality symmetry, with the aim of identifying the bosonic couplings. To achieve this, we propose a truncation scheme for the generalized Buscher rules and the reduced action, specifically targeting the truncation of the nonlinear appearance of the scalar component of the Yang-Mills field in the base space. By imposing this truncated T-duality on the reduced action, we successfully determine the four-derivative bosonic couplings in the minimal basis,  where field redefinition is imposed. Notably, these couplings, which are associated with the Lorentz Chern-Simons coupling $H\Omega$, exhibit an exact correspondence with the NS-NS couplings found in the Metsaev-Tseytlin action.

Furthermore, we investigate the bosonic couplings in the maximal basis, where field redefinition is not imposed. In this scenario, the truncated T-duality fixes the effective action up to 17 arbitrary parameters. By assigning specific values to these parameters, we establish a framework in which the NS-NS couplings align with those in the Meissner action. Remarkably, within this scheme, the Yang-Mills couplings precisely coincide with those obtained through the S-matrix method.
\end{abstract}
\end{titlepage}

\section{Introduction} \label{intro}

It has been proven in string field theory that the classical effective action of string theory, at all orders of derivatives, exhibits a global $O(d,d,\MR)$ symmetry when the effective action is dimensionally reduced on a torus $T^{(d)}$ \cite{Sen:1991zi, Hohm:2014sxa}. In the case of circular reduction, the non-geometric symmetry $O(1,1,\mathbb{Z})$ or the Buscher transformations \cite{Buscher:1987sk,Buscher:1987qj}, together with the gauge symmetries associated with the massless fields and the assumption of the background independence of the effective action at critical dimension \cite{Garousi:2022ovo}, allow for the construction of a covariant and gauge-invariant classical effective action up to overall factors at any order of derivatives.
The Buscher transformations are necessary to incorporate higher derivative corrections \cite{Kaloper:1997ux} that depend on the scheme of the effective action \cite{Garousi:2019wgz}. This approach, as demonstrated in previous works such as \cite{Garousi:2019mca,Garousi:2023kxw,Garousi:2020gio}, has been utilized to derive the Neveu-Schwarz-Neveu-Schwarz (NS-NS) couplings up to eight-derivative orders.
In this paper, we demonstrate that this methodology can also be utilized to determine the Yang-Mills (YM) couplings in the heterotic theory by explicitly performing calculations at the four-derivative order. The four-derivative couplings in the heterotic theory, encompassing both the NS-NS and YM couplings, were previously obtained by studying the sphere-level S-matrix element \cite{Gross:1986mw,Kikuchi:1986rk}.

The two-derivative order bosonic couplings in heterotic theory are given by \cite{Gross:1985rr,Gross:1986mw}
\beqa
{\bf S}^{(0)}&=&-\frac{2}{\kappa^2}\int d^{10}x \sqrt{-G}e^{-2\Phi}\Big[R-\frac{1}{12}H_{\alpha\beta\gamma}H^{\alpha\beta\gamma}+4\nabla_\alpha\Phi\nabla^\alpha\Phi-\frac{1}{4}F_{\mu\nu ij}F^{\mu\nu ij}\Big]\,,\labell{two}
\eeqa
where $\kappa$ is related to the 10-dimensional Newton's constant.
The YM gauge field is defined as $A_{\mu}{}^{ij}=A_{\mu}{}^I(\lambda^I)^{ij}$, where the antisymmetric matrices $(\lambda^I)^{ij}$ represent the adjoint representation of the gauge group $SO(32)$ or $E_8\times E_8$ with the normalization $(\lambda^I)^{ij}(\lambda^J)_{ij}=\delta^{IJ}$. The YM gauge field strength and the $B$-field strength are given by
\beqa
F_{\mu\nu}{}^{ij}&=&\prt_\mu A_{\nu}{}^{ij}-\prt_\nu A_{\mu}{}^{ij}+\frac{1}{\sqrt{\alpha'}}[A_{\mu}{}^{ik},A_{\nu k}{}^{j}]\,\nn\\
H_{\mu\nu\rho}&=&3\prt_{[\mu}B_{\nu\rho]}-\frac{3}{2}A_{[\mu}{}^{ij}F_{\nu\rho]}{}_{ij}\,.\labell{FH}
\eeqa
The NS-NS fields are dimensionless, and we have also normalized the YM gauge field to be dimensionless. Additionally, there is the Lorentz Chern-Simons three-form $\Omega_{\mu\nu\rho}$ in the $B$-field strength, resulting from the Green-Schwarz mechanism \cite{Green:1984sg}. This gives rise to terms with four and six derivatives. It has been explicitly demonstrated in \cite{Hassan:1991mq,  Maharana:1992my,Hohm:2014sxa} that the dimensional reduction of the above action on tori $T^{(d)}$ exhibits global $O(d,d,\mathbb{R})$ symmetry.

We are interested in imposing this symmetry solely to determine the coupling constants in the covariant and gauge-invariant effective actions at any order of derivatives, such as the coefficients $(1, -1/12, 4, -1/4)$ in the action \reef{two}. Therefore, we can impose only the non-geometric aspect of this symmetry on the most general covariant and gauge-invariant couplings, allowing for arbitrary coupling constants. This is because the geometric aspect of this symmetry is already imposed on the effective action through the expression of couplings in covariant and gauge-invariant forms.

Furthermore, the YM gauge symmetry naturally determines the commutator term in the YM field strength. As a result, there is no need to impose the non-geometric part of the $O(d,d,\mathbb{R})$ symmetry specifically to fix the commutator terms in the field strength, allowing us to eliminate these terms from the couplings. It is worth noting that removing the commutator term at the two-derivative order is equivalent to assuming an abelian gauge group. However, in the higher derivative orders of our interest, the matrix indices of the YM gauge field differentiate the couplings from the abelian counterparts.
In this scenario, the  couplings \reef{two}  are limited to two-derivative couplings, and the non-geometric part of $O(d,d,\mathbb{R})$ can determine all the couplings up to a single overall factor. Another simplification arises from considering only the dimensional reduction on a circle, where we can impose the non-geometric group $O(1,1,\mathbb{Z})$ to determine the coupling constants.

The work presented in \cite{Bergshoeff:1995cg} demonstrates the invariance of the aforementioned effective action under $O(1,1,\mathbb{Z})$ transformations. In the proposed reduction scheme for the NS-NS and YM fields in \cite{Maharana:1992my}, these transformations appear as linear transformations for the NS-NS base space fields, while they manifest as nonlinear transformations for the scalar component of the YM field in the base space. This scalar field also appears nonlinearly in the circular reduction of the effective action.

Hence, a further simplification arises by considering the $O(1,1,\mathbb{Z})$ transformations as an expansion involving different powers of the scalar field. In this approach, the reduced action and the $O(1,1,\mathbb{Z})$ transformations can be truncated to include only the zeroth and first orders of the scalar field. Remarkably, we will demonstrate that it is sufficient to consider this truncated reduced action and the $O(1,1,\mathbb{Z})$ transformations to determine the coupling constants. Additionally, it is sufficient to consider the scalar field as constant.
A similar observation regarding the NS-NS couplings in the base space has been made in \cite{Garousi:2019mca}. It has been noted that the $O(1,1,\mathbb{Z})$ constraint imposed on the effective action yields identical relations between the coupling constants, regardless of whether the base space metric is flat or not.  

Once the two-derivative couplings have been determined by imposing the aforementioned truncated T-duality, the subsequent step involves several modifications. First, we substitute the commutator term in the YM field strength, and second, we incorporate the Lorentz Chern-Simons three-form into the $B$-field strength. The inclusion of the Lorentz Chern-Simons three-form introduces additional terms involving four and six derivatives. However, it is worth noting that there are several other couplings at these derivative orders that can be discovered using the same methodology. In the context of this paper, our focus will be on utilizing this approach to determine the couplings at the four-derivative order. We anticipate that this technique can also be applied to uncover higher-derivative couplings in future investigations.

To utilize the aforementioned technique for determining the effective action at higher orders of derivatives, it is necessary to establish a basis for the independent covariant couplings at the given order of derivatives, allowing for arbitrary coupling constants. The coefficients of the general covariant and gauge-invariant couplings at the four-derivative order and beyond are interconnected through integration by parts, the Bianchi identities, and field redefinitions \cite{Metsaev:1987zx}. By employing integration by parts and the Bianchi identities, it becomes possible to reduce the couplings that are solely related through these freedoms to a set of independent couplings. We refer to this as the maximal basis. The coefficients of the couplings in the maximal basis fall into two categories: unambiguous and ambiguous.
Under field redefinitions, the unambiguous parameters remain invariant, while the ambiguous parameters undergo changes. Furthermore, the ambiguous parameters can be divided into essential parameters and arbitrary parameters. The number of essential parameters at each order of derivatives is fixed. The choice of which set of ambiguous parameters is designated as essential parameters determines the basis scheme. If all arbitrary parameters are set to zero, we refer to it as the minimal basis. Other choices for the arbitrary parameters correspond to alternative schemes that are related to the minimal basis through field redefinitions.

In scenarios where the YM field is absent and field redefinitions are involved, the $O(1,1,\mathbb{Z})$ symmetry possesses the capability to determine all parameters in the minimal basis at four derivatives, which consists of 8 couplings \cite{Metsaev:1987zx}, except for a single parameter. In fact, T-duality establishes 7 relations among the 8 unambiguous and essential parameters. On the other hand, if the maximal basis is employed, which encompasses 20 couplings \cite{Metsaev:1987zx}, T-duality  still yields 7 relations among the 20 parameters \cite{Garousi:2022qmk}. In this case, the determination relies on a single unambiguous or essential parameter, accompanied by 12 additional arbitrary parameters. The utilization of field redefinitions enables the elimination of these arbitrary parameters.

In the presence of YM fields added to the NS-NS fields at the four-derivative order, the truncated T-duality establishes 24 relations among the parameters of the independent couplings in both minimal and maximal bases. If we consider the couplings in the minimal basis, which comprises 24 couplings, the truncated T-duality fixes them to be zero. This outcome is expected since there is no string theory wherein the $H$-field possesses a YM Chern-Simons term (see eq.\reef{FH}) but lacks a Lorentz Chern-Simons term. However, when the four-derivative coupling $H\Omega$ is included, obtained through the replacement of the Lorentz Chern-Simons three-form in the $B$-field strength at the two-derivative order, the T-duality determines all 24 parameters in the minimal basis in terms of the coefficient of $H\Omega$.
The NS-NS couplings derived through this approach precisely correspond to those present in the Metsaev-Tseytlin action, which were obtained using the S-matrix method \cite{Metsaev:1987zx}.

If we consider the couplings in the maximal basis, which consists of 42 independent terms, the truncated T-duality still gives rise to 24 relations among the 42 parameters. When the Lorentz Chern-Simons coupling $H\Omega$ is not included, the T-duality fixes the couplings in terms of one unambiguous or essential parameter and 17 additional arbitrary parameters. However, since there is no string theory without a Lorentz Chern-Simons coupling, the condition that all 18 parameters must be removable by field redefinition should be imposed. This condition introduces an extra relation between the parameters.

When the Lorentz Chern-Simons coupling $H\Omega$ is included in the maximal basis, the T-duality generates 24 relations between this term and the 42 couplings. As a result, the couplings can be determined by expressing them in terms of the coefficient of the coupling $H\Omega$ and 18 additional parameters. However, it is crucial to note that the condition of removing these 18 parameters through field redefinition must be imposed, leading to an extra relation between the parameters.
This final step completes the construction of the effective action in the maximal scheme, which involves 17 arbitrary parameters. By assigning specific values to these parameters, the NS-NS couplings can be matched with those in the Meissner action. It is worth mentioning that in the Meissner action, the gravity couplings are in the Gauss-Bonnet scheme, which does not affect the graviton propagator. Furthermore, there are no corrections to the $B$-field and dilaton propagators in this scheme.
Importantly, we have discovered that the YM couplings in this scheme precisely coincide with those found in \cite{Gross:1986mw,Kikuchi:1986rk} through the S-matrix method.

The paper follows the following outline: In Section 2, we investigate the dimensional reduction of the couplings at the two-derivative order to validate the notion that the invariance of the truncated reduced action under the truncated Buscher rules enables the determination of the couplings at this order. Subsequently, we extend this method to the four-derivative couplings in the subsequent sections. 
In Section 3, we identify the independent couplings at the four-derivative order. In Subsection 3.1, we determine the couplings in the minimal basis, where we enforce field redefinition as well as integration by parts and Bianchi identities to reduce the number of independent couplings to 24. In Subsection 3.2, we forgo the field redefinition and instead focus on removing redundancy through integration by parts and Bianchi identities to obtain the set of 42 independent couplings.
Moving on to Section 4, we impose truncated T-duality on these bases to fix the coupling constants. In Subsection 4.1, we fix the couplings in the minimal basis and find that there are 10 couplings. Notably, four of these couplings, which solely involve the NS-NS field, match exactly with the couplings in the Metsaev-Tseytlin action. The YM couplings in this scheme exhibit variations from those obtained in the literature through the S-matrix method, owing to the fact that the graviton propagators receive four-derivative corrections in the Metsaev-Tseytlin action. 
In Subsection 4.2, we fix the couplings in the maximal basis using the  truncated T-duality. In this case, the effective action comprises 17 arbitrary parameters. By assigning specific values to these parameters, the NS-NS couplings align with those in the Meissner action, where the propagators do not receive four-derivative corrections. Remarkably, we demonstrate that the YM couplings precisely match those obtained through the S-matrix method.
In Section 5, we provide a brief discussion of our findings. Throughout our calculations, we employ the "xAct" package \cite{Nutma:2013zea}.

\section{Effective Action at the Two-Derivative Order}\label{sec.2}

It is well-established that the two-derivative effective action of heterotic theory, when dimensionally reduced on a torus $T^{(d)}$, exhibits invariance under $O(d,d,\MR)$ transformations \cite{Hassan:1991mq,  Maharana:1992my,Hohm:2014sxa}. The circular invariance of this action, governed by the $\MZ_2$-Buscher rules, has been extensively studied in \cite{Bergshoeff:1995cg}. Notably, the scalar component of the YM gauge field appears nonlinearly in both the reduced action and the Buscher rules \cite{Bergshoeff:1995cg}.

In this context, we leverage the aforementioned symmetry to determine the coupling constants of the covariant and gauge invariant couplings. To achieve this, we adopt a power expansion for the Buscher rules and subsequently truncate terms in the power expansion of both the Buscher rules and the reduced effective action, specifically excluding those involving more than one YM scalar field. Remarkably, we will demonstrate that such a truncation provides sufficient information to fix the effective action up to an overall factor.

The most general covariant and gauge-invariant independent coupling involving the NS-NS and YM fields at the two-derivative order is given by the following action:
\beqa
{\bf S}^{(0)}&=&-\frac{2}{\kappa^2}\int d^{10}x \sqrt{-G}e^{-2\Phi}\Big[a_1R-\frac{a_3}{12}H_{\alpha\beta\gamma}H^{\alpha\beta\gamma}+4a_2\nabla_\alpha\Phi\nabla^\alpha\Phi-\frac{a_4}{4}F_{\mu\nu ij}F^{\mu\nu ij}\Big]\,.\labell{two2}
\eeqa
 The YM gauge field strength and the $B$-field strength are given by equation \reef{FH}, neglecting the commutator term. The coupling constants $a_1, \ldots, a_4$ are to be determined by imposing the truncated Bucher rules.

To perform the dimensional reduction of the action on a circle, we employ the following reduction scheme for the NS-NS and YM fields \cite{Maharana:1992my,Bergshoeff:1995cg}:
 \beqa
G_{\mu\nu}=\left(\matrix{\bg_{ab}+e^{\varphi}g_{a }g_{b }& e^{\varphi}g_{a }&\cr e^{\varphi}g_{b }&e^{\varphi}&}\!\!\!\!\!\right),\, \, A_{\mu}{}^{ij}=\left(\matrix{\bar{A}_a{}^{ij}+e^{\varphi/2}\alpha^{ij}g_a&\cr e^{\varphi/2}\alpha^{ij}&}\!\!\!\!\!\right),\, \Phi=\bar{\phi}+\varphi/4\,,\labell{reduc}
\eeqa
\beqa
B_{\mu\nu}= \left(\matrix{\bb_{ab}+\frac{1}{2}(b_{a }g_{b }-b_{b }g_{a })+\frac{1}{2}e^{\varphi/2}\alpha_{ij}(g_a\bar{A}_b{}^{ij}-g_b\bar{A}_a{}^{ij})&b_{a }-\frac{1}{2}e^{\varphi/2}\alpha_{ij}\bar{A}_a{}^{ij}\cr - b_{b }+\frac{1}{2}e^{\varphi/2}\alpha_{ij}\bar{A}_a{}^{ij}&0&}\!\!\!\!\!\right),\nn
\eeqa
where $\bar{g}_{ab}$, $\bar{b}_{ab}$, $\bar{\phi}$, and $\bar{A}_a^{ij}$ represent the metric, B-field, dilaton, and YM gauge field in the base space, respectively. Additionally, $g_a$ and $b_b$ denote two vectors, while $\varphi$ and $\alpha^{ij}$ represent scalars within this space. The difference between the above reduction and those in \cite{Bergshoeff:1995cg} is that we include a factor of $e^{\varphi/2}$ to the scalar component of the YM gauge field. By employing the above reduction scheme, one can derive the following reductions for different components of the $H$-field \cite{Maharana:1992my,Bergshoeff:1995cg}:
\beqa
H_{aby}&=&W_{ab}-\frac{1}{2}e^{\vp}\alpha^2 V_{ab}-e^{\vp/2}\alpha^{ij}\bF_{ab ij}\,,\nn\\
H_{abc}&=&\bH_{abc}+3g_{[a}W_{bc]}-\frac{3}{2}e^{\vp}\alpha^2g_{[a}V_{bc]}-3e^{\vp/2}\alpha_{ij}g_{[a}\bF_{bc]}{}^{ij}\,,
\eeqa
where $\alpha^2=\alpha^{ij}\alpha_{ij}$, $W_{ab}=\partial_a b_b-\partial_b b_a$, $V_{ab}=\partial_a g_b-\partial_b g_a$, $\bar{F}_{ab}{}^{ij}=\partial_a\bar{A}_a{}^{ij}-\partial_b\bar{A}_a{}^{ij}$, and $\bar{H}$ represents the torsion in the base space:
\beqa
\bH_{abc}&=&3\prt_{[a}\bb_{bc]}-\frac{3}{2}g_{[a}W_{bc]}-\frac{3}{2}b_{[a}V_{bc]}-\frac{3}{2}\bA_{[a}{}^{ij}\bF_{bc]}{}^{ij}\,.
\eeqa
Our notation for antisymmetry is such that, for example, $3g_{[a}W_{bc]}=g_aW_{bc}-g_bW_{ac}-g_cW_{ba}$.

The circular reduction of the action \reef{two2} can be expressed in terms of the field strengths $W$, $V$, $\bar{F}$, and $\bar{H}$ in the base space as follows:
\beqa
S^{(0)}&\!\!\!\!\!=\!\!\!\!\!&-\frac{2}{\kappa'^2}\int d^{9}x \sqrt{-\bg}\,e^{-2\bphi}\Big[a_1\bar{R}-\frac{a_3}{12}\bH^2+4a_2\prt_a\bphi\prt^a\bphi-\frac{a_4}{4}\bF_{ab ij}\bF^{ab ij}+2(a_2-a_1)\prt_a\vp\prt^a\bphi\nn\\
&&+\frac{1}{4}(a_2-2a_1)\prt_a\vp\prt^a\vp-\frac{a_1}{4}e^{\vp}V^2-\frac{a_3}{4}e^{-\vp}W^2-\frac{a_4}{2}e^{\vp/2}\bF_{ab ij}V^{ab}\alpha^{ij}+\frac{a_3}{2}e^{-\vp/2}\bF_{ab ij}W^{ab}\alpha^{ij}\nn\\
&&-\frac{a_4}{2}\prt_a\alpha^{ij}\prt^a\alpha_{ij}+\alpha^2\Big(-\frac{a_4}{8}\prt_a\vp\prt^a\vp-\frac{a_4}{4}e^{\vp}V^2+\frac{a_3}{4}V^{ab}W_{ab}-\frac{a_3}{4}\bF_{ab ij}\bF^{abij}\nn\\&&-\frac{a_3}{4}e^{\vp/2}\bF_{ab ij} V^{ab}\alpha^{ij}-\frac{a_3}{16}e^{\vp}V^2\alpha^2\Big)\Big]\,,\labell{two3}
\eeqa
where $\kappa'$ is related to the 9-dimensional Newton's constant. In deriving this expression, we have employed integration by parts. The terms in the third and fourth lines exhibit nonlinearity with respect to YM scalar field $\alpha^{ij}$, whereas the terms in the first and second lines are of zeroth and first order in $\alpha^{ij}$. Additionally, all coupling constants $a_1,\cdots, a_4$ appear in the first and second lines. 

Now, one can observe that the terms in the first and second lines remain invariant under the following linear transformations:
\beqa
 g_{a}\rightarrow b_{a}, b_{a}\rightarrow g_{a}, \vp\rightarrow -\vp, \alpha^{ij}\rightarrow -\alpha^{ij},\bA_a{}^{ij}\rightarrow \bA_a{}^{ij}, \bg_{ab}\rightarrow \bg_{ab},  \bb_{ab}\rightarrow \bb_{ab},\bphi\rightarrow \bphi\,,\labell{bucher}
 \eeqa
provided that the parameters in the action are all the same, i.e., $a_1=a_2=a_3=a_4$. One can extend the aforementioned linear transformation to a nonlinear transformation and observe that the terms in the third and fourth lines of \reef{two3} remain invariant under the nonlinear Buscher transformations \cite{Bergshoeff:1995cg}. However, such calculations serve only as a demonstration that the full action is indeed invariant under the nonlinear Buscher transformation, which is not our primary focus. Therefore, to utilize the $O(1,1,\mathbb{Z})$ symmetry in determining the coupling constants, it suffices to consider the terms of zeroth and first order in $\alpha^{ij}$ and treat these scalars as constants in the reduced action and Buscher rules. We speculate that this truncation in the reduced action and Buscher rules could potentially fix the couplings even at higher orders of derivatives. In the subsequent sections, we will perform this calculation for the couplings involving four derivatives.

\section{Bases for Four-Derivative Couplings}

To determine the effective action using the T-duality $\mathbb{Z}_2$-constraint, it is necessary to first identify a basis with unfixed coupling constants. Subsequently, the $\mathbb{Z}_2$-symmetry needs to be imposed on this basis in order to establish the relationships between the coupling constants. 
For the identification of the basis at the four-derivative order, as outlined in \cite{Garousi:2019cdn}, it is essential to list all covariant and gauge-invariant NS-NS and YM couplings. By utilizing the "xAct" package \cite{Nutma:2013zea}, it has been determined that there are a total of 81 such couplings. None of them can have an odd number of YM field strengths. These couplings can be expressed as follows:
\beqa
\mathcal{L}'_1=& c'_1 F_\alpha{}^{\mu ij}F^{\alpha\beta}{}_{ ij}F_\beta{}^{\nu kl}F_{\mu\nu}{}_{kl}+c'_2 F_\alpha{}^{\mu ij}F^{\alpha\beta}{}_{ il}F_\beta{}^{\nu}{}_{ jk}F_{\mu\nu}{}^{kl}+\cdots\,.\labell{L3}
\eeqa
Here, $c'_1, \ldots, c'_{81}$ denote the coupling constants. However,  these  couplings are not all independent. Some of them are interrelated through total derivative terms, while others are connected through field redefinitions or various Bianchi identities.

To eliminate the total derivative terms from the aforementioned couplings, the following terms are introduced into the Lagrangian $\mathcal{L}'_1$:
\beqa
e^{-2\Phi} \mathcal{J}_1&\equiv&  \nabla_\alpha (e^{-2\Phi}{\cal I}_1^\alpha)\,.\labell{J3}
\eeqa
Here, the vector ${\cal I}_1^\alpha$ represents a collection of all possible covariant and gauge-invariant terms at the three-derivative level, including arbitrary parameters.

To eliminate the freedom of field redefinitions, it is necessary to perturb the metric, dilaton, $H$-field, and YM gauge fields in the two-derivative action described in \reef{two}. By utilizing the Bianchi identity satisfied by the $H$-field:
\beqa
\nabla_{[\alpha}H_{\beta\mu\nu]}+\frac{3}{4}F_{[\alpha\beta}{}^{ij}F_{\mu\nu]}{}_{ij}&=&0\,,\labell{iden0}
\eeqa
it is found that the perturbation of the $H$-field and the perturbation of the YM field are related through the equation $d(\delta H+3F^{ij}\delta A^{ij})=0$, where form notation is employed. Consequently, the perturbation of the $H$-field can be expressed as a linear combination of the YM gauge field perturbation and an arbitrary two-form $\delta B$:
\beqa
\delta H&=&3d\delta B-3F^{ij}\delta A_{ij}\,.
\eeqa 
Subsequently, the field redefinition introduces the following terms into the Lagrangian $\mathcal{L}'_1$:
\beqa
\mathcal{K}_1
&\!\!\!\!\!\equiv\!\!\!\!\!\!& (\frac{1}{2} \nabla_{\gamma}H^{\alpha \beta \gamma} -  H^{\alpha \beta}{}_{\gamma} \nabla^{\gamma}\Phi)\delta B_{\alpha\beta}\nn\\
&&-(\nabla^{\beta}F_{\alpha\beta}{}^{ij}-2F_{\alpha\beta}{}^{ij}\nabla^{\beta}\Phi-\frac{1}{2}F^{\beta\mu ij} H_{\alpha\beta\mu})\delta A^{\alpha}{}_{ ij}\nn\\
&& -(  R^{\alpha \beta}-\frac{1}{4} H^{\alpha \gamma \delta} H^{\beta}{}_{\gamma \delta}+ 2 \nabla^{\beta}\nabla^{\alpha}\Phi-\frac{1}{2}F^{\alpha\mu ij} F^{\beta}{}_{\mu ij})\delta G_{\alpha\beta}\labell{eq.13}\\
\nn\\
&&-2( R\! -\!\frac{1}{12} H_{\alpha \beta \gamma} H^{\alpha \beta \gamma} \!+\! 4 \nabla_{\alpha}\nabla^{\alpha}\Phi \!-4 \nabla_{\alpha}\Phi \nabla^{\alpha}\Phi-\frac{1}{4}F_{\alpha\beta ij}F^{\alpha\beta ij}) (\delta\Phi-\frac{1}{4}\delta G^{\mu}{}_\mu) .\nn
\eeqa
In this expression, the perturbations $\delta G_{\mu\nu}, \delta B_{\mu\nu},\delta \Phi, \delta A_a{}^{ij}$ are constructed from the NS-NS and YM fields at the two-derivative order, with arbitrary coefficients. $\delta G_{\mu\nu}$ is symmetric, and $\delta B_{\mu\nu}$ is antisymmetric. By incorporating the total derivative terms and the contribution from field redefinitions into the Lagrangian $\mathcal{L}'_1$, the resulting Lagrangian, denoted as $\mathcal{L}_1$, exhibits different coupling constants. 
Consequently, the equation
\beqa
\Delta_1-{\cal J}_1-{\cal K}_1&=&0\,,\labell{DLK}
\eeqa
holds, where $\Delta_1 = \mathcal{L}_1 - \mathcal{L}'_1$ is equivalent to $\mathcal{L}'_1$, but with coefficients $\delta c_1, \delta c_2, \ldots$, where $\delta c_i = c_i - c'_i$. Solving the above equation yields linear relations among $\delta c_1, \delta c_2, \ldots$, which indicate how the couplings are related to each other through total derivative and field redefinition terms. Additionally, there are relations between $\delta c_1, \delta c_2, \ldots$ and the coefficients of the total derivative terms and field redefinitions, but they are not of interest for our purposes.

To solve the equation \reef{DLK}, it is necessary to express it in terms of independent couplings by imposing the following Bianchi identities:
\beqa
 R_{\alpha[\beta\gamma\delta]}=0\,,&&
 \nabla_{[\mu}R_{\alpha\beta]\gamma\delta}=0\,,\,\,\,\,\,\,
{[}\nabla,\nabla{]}\mathcal{O}-R\mathcal{O}=0\,,\nn\\
\nabla_{[\alpha}H_{\beta\mu\nu]}+\frac{3}{4}F_{[\alpha\beta}{}^{ij}F_{\mu\nu]}{}_{ij}=0\,,&&
\nabla_{[\alpha}F_{\beta\gamma]}{}^{ij}=0\,.\labell{bian}
\eeqa
If we had not removed the commutator term in the YM field strength, an additional Bianchi identity would arise as $[\nabla, \nabla]F\sim FF$. This relation connects the coupling that involves antisymmetric two-derivatives of $F$ to the coupling that involves $FF$. In other words, these two couplings are not independent. If one chooses one of them as an independent coupling, the other one must be eliminated from the list of independent couplings. By removing the commutator term in the YM field strength, we consider the couplings $FF$ to be independent and remove the couplings that involve antisymmetric two-derivatives of the YM field strength from the list of all higher derivative couplings.

The Bianchi identities \reef{bian} can be imposed on the equation \reef{DLK} either in a gauge-invariant form or a non-gauge-invariant form.  We impose them  in a non-gauge-invariant form. Hence, we rewrite the terms in \reef{DLK} in a local frame where the covariant derivatives are expressed in terms of partial derivatives and the first partial derivative of the metric is zero. Additionally, the terms involving $H$ and $F$ in \reef{DLK} can be rewritten in terms of potentials using the relations \reef{FH}, in which the commutator is removed. This way, all the Bianchi identities are automatically satisfied \cite{Garousi:2019cdn}.

The equation \reef{DLK} can then be solved to find the basis. We consider solving this equation in two cases: using field redefinition or not. In the next subsection, we use field redefinition to find the couplings in the minimal basis. In the subsequent subsection, we do not use field redefinition, i.e., we do not include ${\cal K}_1$ in this equation, to find the couplings in the maximal basis.

\subsection{ Minimal Basis}

Imposing the Bianchi identity in a non-gauge-invariant form, one can rewrite the different terms on the left-hand side of \reef{DLK} in terms of independent but non-gauge-invariant couplings. The coefficients of the independent terms must be zero, leading to a set of algebraic equations. The solution to these equations has two parts. The first part consists of 24 relations between only    $\delta c_i$'s, while the second part consists of additional relations between the coefficients of total derivative terms, field redefinitions, and $\delta c_i$'s, which are not of interest to us. The number of relations in the first part determines the number of independent couplings in ${\cal L}_1'$. In a particular scheme, it is possible to set some of the coefficients in ${\cal L}_1'$ to zero. However, after replacing the non-zero terms in \reef{DLK}, the number of relations between only $\delta c_i$'s should remain unchanged. In other words, there must always be 24 relations.

For the terms involving only NS-NS couplings, we include the 8 independent terms from the minimal basis of NS-NS fields chosen \cite{Metsaev:1987zx} by setting all other NS-NS terms to zero. As for the YM couplings, we set the couplings involving derivatives of $H$ and $F$ to zero, along with those involving the Ricci tensor, Ricci scalar, and couplings with an odd number of $H$-fields. Despite eliminating these terms, equation \reef{DLK} still yields more than 24 $\delta c_i$ in the relations that solely involve $\delta c_i$. This suggests that we can further set some of the $\delta c_i$ to zero to obtain a set of 24 relations characterized by $\delta c_i = 0$.

Among the couplings involving YM field strengths, there are 8 terms that feature four field strengths. Four of these couplings involve single-trace terms in the gauge group indices, such as $\Tr(F_{\mu\nu}F^{\mu\nu} F_{\alpha\beta}F^{\alpha\beta})$, while the other four involve two-trace terms, such as $\Tr(F_{\mu\nu}F^{\mu\nu})\Tr(F_{\alpha\beta}F^{\alpha\beta})$. The coupling constants associated with the single-trace terms are unambiguous parameters, whereas the coupling constants of the two-trace terms are ambiguous parameters. In our chosen scheme, we consider the two-trace terms as independent couplings. There are still other choices available for selecting the independent couplings.

We select the 24 couplings according to the following scheme:
\beqa
\mathcal{L}_1&=&\alpha'\Big[c_{1}  
    F_{\alpha  }{}^{\gamma  kl} F^{\alpha  \beta  ij} 
F_{\beta  }{}^{\delta  }{}_{kl} F_{\gamma  \delta  ij} +   
   c_{2}  
    F_{\alpha  }{}^{\gamma  }{}_{i}{}^{k} F^{\alpha  \beta  
ij} F_{\beta  }{}^{\delta  }{}_{k}{}^{l} F_{\gamma  \delta  jl} 
+   c_{3}  
    F_{\alpha  }{}^{\gamma  }{}_{i}{}^{k} F^{\alpha  \beta  
ij} F_{\beta  }{}^{\delta  }{}_{j}{}^{l} F_{\gamma  \delta  kl} \nn\\&&
+   c_{4}  
    F_{\alpha  }{}^{\gamma  }{}_{ij} F^{\alpha  \beta  ij} 
F_{\beta  }{}^{\delta  kl} F_{\gamma  \delta  kl} +   
   c_{5}  
    F_{\alpha  \beta  }{}^{kl} F^{\alpha  \beta  ij} 
F_{\gamma  \delta  kl} F^{\gamma  \delta  }{}_{ij} +   
   c_{6}  
    F_{\alpha  \beta  }{}^{kl} F^{\alpha  \beta  ij} 
F_{\gamma  \delta  jl} F^{\gamma  \delta  }{}_{ik}  \nn\\&&+   
   c_{7}  
    F_{\alpha  \beta  i}{}^{k} F^{\alpha  \beta  ij} 
F_{\gamma  \delta  kl} F^{\gamma  \delta  }{}_{j}{}^{l} +   
   c_{8}  
    F_{\alpha  \beta  ij} F^{\alpha  \beta  ij} F_{\gamma  
\delta  kl} F^{\gamma  \delta  kl} +   c_{9}  
    F^{\alpha  \beta  ij} F^{\gamma  \delta  }{}_{ij} 
H_{\alpha  \gamma  }{}^{\epsilon  } H_{\beta  \delta  \epsilon  
}  \nn\\&&+   c_{10}  
    F^{\alpha  \beta  ij} F^{\gamma  \delta  }{}_{ij} 
H_{\alpha  \beta  }{}^{\epsilon  } H_{\gamma  \delta  \epsilon  
} +   c_{11}  
    F_{\alpha  }{}^{\gamma  }{}_{ij} F^{\alpha  \beta  ij} H_{
\beta  }{}^{\delta  \epsilon  } H_{\gamma  \delta  \epsilon  } 
+   c_{12}  
    H_{\alpha  }{}^{\delta  \epsilon  } H^{\alpha  \beta  
\gamma  } H_{\beta  \delta  }{}^{\varepsilon  } H_{\gamma  
\epsilon  \varepsilon  }  \nn\\&&+   c_{13}  
    F_{\alpha  \beta  ij} F^{\alpha  \beta  ij} H_{\gamma  
\delta  \epsilon  } H^{\gamma  \delta  \epsilon  } +   
   c_{14}  
    H_{\alpha  \beta  }{}^{\delta  } H^{\alpha  \beta  \gamma  
} H_{\gamma  }{}^{\epsilon  \varepsilon  } H_{\delta  \epsilon  
\varepsilon  } +   c_{15}  
    H_{\alpha  \beta  \gamma  } H^{\alpha  \beta  \gamma  } 
H_{\delta  \epsilon  \varepsilon  } H^{\delta  \epsilon  
\varepsilon  } \nn\\&& +   c_{16}  
    F^{\alpha  \beta  ij} F^{\gamma  \delta  }{}_{ij} 
R_{\alpha  \beta  \gamma  \delta  } +   c_{17}  
    R_{\alpha  \beta  \gamma  \delta  } 
R^{\alpha  \beta  \gamma  \delta  } +   c_{18}  
    H_{\alpha  }{}^{\delta  \epsilon  } H^{\alpha  \beta  
\gamma  } R_{\beta  \gamma  \delta  \epsilon  }\labell{min} \\&& +   
   c_{19}  
    F_{\beta  \gamma  ij} F^{\beta  \gamma  ij} 
\nabla_{\alpha  }\Phi  \nabla^{\alpha  }\Phi  +   
   c_{20}  
    H_{\beta  \gamma  \delta  } H^{\beta  \gamma  \delta  } 
\nabla_{\alpha  }\Phi  \nabla^{\alpha  }\Phi  +   
   c_{21}  
    H_{\alpha  }{}^{\gamma  \delta  } H_{\beta  \gamma  \delta 
 } \nabla^{\alpha  }\Phi  \nabla^{\beta  }\Phi  \nn\\&& \!+\!   
   c_{22}  
    \nabla_{\alpha  }\Phi  \nabla^{\alpha  }\Phi  
\nabla_{\beta  }\Phi  \nabla^{\beta  }\Phi  \!+\!   c_{23} 
     F_{\alpha  }{}^{\beta  ij} \nabla^{\alpha  }\Phi  
\nabla_{\gamma  }F_{\beta  }{}^{\gamma  }{}_{ij}\!+\!c_{24}F_\alpha{}^{\beta ij}\nabla^\alpha\Phi\nabla_\gamma F_\beta{}^\gamma{}_{ij}\!+\!\frac{1}{4}H^{\alpha\beta\gamma}\Omega_{\alpha\beta\gamma}\Big],\nn
\eeqa
where we have also included the four-derivative $H\Omega$ term, which results from replacing the Lorentz Chern-Simons three-form in the $H$-field \cite{Green:1984sg}, i.e., $H\rightarrow H-\frac{3\alpha'}{2}\Omega$, in the leading-order action \reef{two}. The Chern-Simons three-form is given by:
\beqa
\Omega_{\mu\nu\alpha}&=&\omega_{[\mu {\mu_1}}{}^{\nu_1}\prt_\nu\omega_{\alpha] {\nu_1}}{}^{\mu_1}+\frac{2}{3}\omega_{[\mu {\mu_1}}{}^{\nu_1}\omega_{\nu {\nu_1}}{}^{\alpha_1}\omega_{\alpha]{\alpha_1}}{}^{\mu_1}\,\,;\,\,\,\omega_{\mu {\mu_1}}{}^{\nu_1}=e^\nu{}_{\mu_1}\nabla_\mu e_\nu{}^{\nu_1} \,,
\eeqa
where $e_\mu{}^{\mu_1}e_\nu{}^{\nu_1}\eta_{\mu_1\nu_1}=G_{\mu\nu}$. Our index convention is that $\mu, \nu, \ldots$ are the indices of the curved spacetime, and $\mu_1, \nu_1, \ldots$ are the indices of the flat tangent spaces. We assume that the effective action has an overall factor of $-2/\kappa^2$.

The parameters $c_1, \ldots, c_{24}$ in the above Lagrangian are background-independent coupling constants that can be determined either through the S-matrix method, which involves comparing the S-matrix of the field theory with the S-matrix elements in string theory in flat spacetime, or through T-duality. In the T-duality approach, one direction of spacetime is assumed to be a circle, and the truncated $O(1,1,\mathbb{Z})$ symmetry is imposed on the coupling constants. We employ the latter approach to determine the coupling constants in the minimal basis.

\subsection{Maximal Basis}

To obtain the maximal basis, one should refrain from using field redefinition. By removing ${\cal K}_1$ from the equation \reef{DLK} and solving it again, one discovers 42 relations solely between $\delta c_i$'s. Consequently, the maximal basis consists of 42 couplings. In this scenario, it is permissible to set the coefficients of the couplings involving first derivatives of curvatures, second derivatives of $H$ and $F$, and third derivatives of the dilaton to zero. After setting them to zero, the equation \reef{DLK} still generates 42 relations solely between $\delta c_i$'s. However, there are still more than 42 $\delta c_i$  present in these relations, indicating that we are still able to set some of them to zero in order to produce 42 relations $\delta c_i=0$. Unlike in the minimal basis, we are not allowed to set all terms with an odd number of $H$ to zero.

We choose the 42 couplings in the following scheme:
\beqa
\mathcal{L}_1&=&\alpha'\Big[c_{1}  
    F_{\alpha  }{}^{\gamma  kl} F^{\alpha  \beta  ij} 
F_{\beta  }{}^{\delta  }{}_{kl} F_{\gamma  \delta  ij} +   
   c_{2}  
    F_{\alpha  }{}^{\gamma  }{}_{i}{}^{k} F^{\alpha  \beta  
ij} F_{\beta  }{}^{\delta  }{}_{k}{}^{l} F_{\gamma  \delta  jl} 
+   c_{3}  
    F_{\alpha  }{}^{\gamma  }{}_{i}{}^{k} F^{\alpha  \beta  
ij} F_{\beta  }{}^{\delta  }{}_{j}{}^{l} F_{\gamma  \delta  kl} \nn\\&&
+   c_{4}  
    F_{\alpha  }{}^{\gamma  }{}_{ij} F^{\alpha  \beta  ij} 
F_{\beta  }{}^{\delta  kl} F_{\gamma  \delta  kl} +   
   c_{5}  
    F_{\alpha  \beta  }{}^{kl} F^{\alpha  \beta  ij} 
F_{\gamma  \delta  kl} F^{\gamma  \delta  }{}_{ij} +   
   c_{6}  
    F_{\alpha  \beta  }{}^{kl} F^{\alpha  \beta  ij} 
F_{\gamma  \delta  jl} F^{\gamma  \delta  }{}_{ik}\nn\\&& +   
   c_{7}  
    F_{\alpha  \beta  i}{}^{k} F^{\alpha  \beta  ij} 
F_{\gamma  \delta  kl} F^{\gamma  \delta  }{}_{j}{}^{l} +   
   c_{8}  
    F_{\alpha  \beta  ij} F^{\alpha  \beta  ij} F_{\gamma  
\delta  kl} F^{\gamma  \delta  kl} +   c_{9}  
    F^{\alpha  \beta  ij} F^{\gamma  \delta  }{}_{ij} 
H_{\alpha  \gamma  }{}^{\epsilon  } H_{\beta  \delta  \epsilon  
}\nn\\&& +   c_{10}  
    F^{\alpha  \beta  ij} F^{\gamma  \delta  }{}_{ij} 
H_{\alpha  \beta  }{}^{\epsilon  } H_{\gamma  \delta  \epsilon  
} +   c_{11}  
    F_{\alpha  }{}^{\gamma  }{}_{ij} F^{\alpha  \beta  ij} H_{
\beta  }{}^{\delta  \epsilon  } H_{\gamma  \delta  \epsilon  } 
+   c_{12}  
    H_{\alpha  }{}^{\delta  \epsilon  } H^{\alpha  \beta  
\gamma  } H_{\beta  \delta  }{}^{\varepsilon  } H_{\gamma  
\epsilon  \varepsilon  } \nn\\&&+   c_{13}  
    F_{\alpha  \beta  ij} F^{\alpha  \beta  ij} H_{\gamma  
\delta  \epsilon  } H^{\gamma  \delta  \epsilon  } +   
   c_{14}  
    H_{\alpha  \beta  }{}^{\delta  } H^{\alpha  \beta  \gamma  
} H_{\gamma  }{}^{\epsilon  \varepsilon  } H_{\delta  \epsilon  
\varepsilon  } +   c_{15}  
    H_{\alpha  \beta  \gamma  } H^{\alpha  \beta  \gamma  } 
H_{\delta  \epsilon  \varepsilon  } H^{\delta  \epsilon  
\varepsilon  } \nn\\&&+   c_{16}  
    F_{\alpha  }{}^{\gamma  ij} F_{\beta  \gamma  ij} 
R^{\alpha  \beta  } +   c_{17}  
    H_{\alpha  }{}^{\gamma  \delta  } H_{\beta  \gamma  \delta 
 } R^{\alpha  \beta  } +   c_{18}  
    R_{\alpha  \beta  } R^{\alpha  \beta  } + 
  c_{19}  
    F_{\alpha  \beta  ij} F^{\alpha  \beta  ij} R\nn\\&& +  
 c_{20}  
    H_{\alpha  \beta  \gamma  } H^{\alpha  \beta  \gamma  } R +   c_{21}   R^2 +   c_{22}  
    F^{\alpha  \beta  ij} F^{\gamma  \delta  }{}_{ij} 
R_{\alpha  \beta  \gamma  \delta  } +   c_{23}  
    R_{\alpha  \beta  \gamma  \delta  } 
R^{\alpha  \beta  \gamma  \delta  } +   c_{24}  
    H_{\alpha  }{}^{\delta  \epsilon  } H^{\alpha  \beta  
\gamma  } R_{\beta  \gamma  \delta  \epsilon  } \nn\\&&+   
   c_{25}  
    F^{\alpha  \beta  ij} H_{\beta  \gamma  \delta  } 
\nabla_{\alpha  }F^{\gamma  \delta  }{}_{ij} +   c_{26} 
     F_{\beta  \gamma  ij} F^{\beta  \gamma  ij} 
\nabla_{\alpha  }\nabla^{\alpha  }\Phi  +   c_{27}  
    H_{\beta  \gamma  \delta  } H^{\beta  \gamma  \delta  } 
\nabla_{\alpha  }\nabla^{\alpha  }\Phi  +   c_{28}  
    R \nabla_{\alpha  }\nabla^{\alpha  }\Phi\nn\\&&  +   
   c_{29}  
    F_{\beta  \gamma  ij} F^{\beta  \gamma  ij} 
\nabla_{\alpha  }\Phi  \nabla^{\alpha  }\Phi  +   
   c_{30}  
    H_{\beta  \gamma  \delta  } H^{\beta  \gamma  \delta  } 
\nabla_{\alpha  }\Phi  \nabla^{\alpha  }\Phi \! +\!   
   c_{31}  
    R \nabla_{\alpha  }\Phi  \nabla^{\alpha  }\Phi \! + \!
  c_{32}  
    \nabla_{\alpha  }\nabla^{\alpha  }\Phi  \nabla_{\beta  
}\nabla^{\beta  }\Phi  \nn\\&&+   c_{33}  
    F_{\alpha  }{}^{\gamma  ij} F_{\beta  \gamma  ij} 
\nabla^{\alpha  }\Phi  \nabla^{\beta  }\Phi  +   c_{34} 
     H_{\alpha  }{}^{\gamma  \delta  } H_{\beta  \gamma  
\delta  } \nabla^{\alpha  }\Phi  \nabla^{\beta  }\Phi  +   
   c_{35}  
    R_{\alpha  \beta  } \nabla^{\alpha  }\Phi  
\nabla^{\beta  }\Phi \nn\\&& +   c_{36}  
    \nabla_{\alpha  }\Phi  \nabla^{\alpha  }\Phi  
\nabla_{\beta  }\Phi  \nabla^{\beta  }\Phi  +   c_{37} 
     \nabla^{\alpha  }\Phi  \nabla_{\beta  }\nabla_{\alpha  }
\Phi  \nabla^{\beta  }\Phi  +   c_{38}  
    F_{\alpha  }{}^{\gamma  ij} F_{\beta  \gamma  ij} 
\nabla^{\beta  }\nabla^{\alpha  }\Phi  \nn\\&&+   c_{39}  
    H_{\alpha  }{}^{\gamma  \delta  } H_{\beta  \gamma  \delta 
 } \nabla^{\beta  }\nabla^{\alpha  }\Phi  +   c_{40}  
   \nabla_{\beta  }F_{\alpha  \gamma  ij} 
\nabla^{\gamma  }F^{\alpha  \beta  }{}^{ij} +   c_{41}  
    H^{\beta  \gamma  \delta  } \nabla^{\alpha  }\Phi  
\nabla_{\delta  }H_{\alpha  \beta  \gamma  } \nn\\&&+   c_{42} 
     \nabla_{\gamma  }H_{\alpha  \beta  \delta  } 
\nabla^{\delta  }H^{\alpha  \beta  \gamma  }+\frac{1}{4}H^{\alpha\beta\gamma}\Omega_{\alpha\beta\gamma}\Big]\,,\labell{max}
\eeqa
where we have also included the four-derivative $H\Omega$ term. It is worth noting that if the YM gauge field is set to zero, the basis consists of 21 couplings, whereas if one sets it to zero before solving the equation \reef{DLK}, one would find 20 couplings. This discrepancy is not an inconsistency because in the presence of the YM field, the Bianchi identity \reef{iden0} can relate terms without $F$ to terms that include $F$, or terms with $F$ to terms without $F$. It is possible to find the maximal basis in a different scheme such that if the YM field is removed, it becomes the same as the maximal basis for only NS-NS fields. In fact, in the minimal basis \reef{min}, we have chosen such a scheme.

The parameters $c_1, \ldots, c_{42}$ in the above Lagrangian are background-independent coupling constants that we will determine in the next section by imposing the truncated T-duality.

\section{Effective Action at the Four-Derivative Order}

The observation that the dimensional reduction of the classical effective action of string theory on a torus $T^{(d)}$ must be invariant under the $O(d,d, \mathbb{R})$ transformations \cite{Sen:1991zi,Hohm:2014sxa} indicates that the circular reduction of the couplings in the effective action should be invariant under the discrete group $O(1,1, \mathbb{Z})$ or $\mathbb{Z}_2$-group, which consists only of non-geometrical transformations. Hence, to impose this $\mathbb{Z}_2$-constraint on the classical effective action ${\bf S}_{\rm eff}$, one needs to reduce the theory on a circle using the reduction \reef{reduc} to obtain the $(D-1)$-dimensional effective action $S_{\rm eff}(\psi)$ where $\psi$ collectively represents the base space fields. Then, one transforms this action under the $\mathbb{Z}_2$-transformations to produce $S_{\rm eff}(\psi')$ where $\psi'$ is the $\mathbb{Z}_2$-transformation of $\psi$. The $\mathbb{Z}_2$-constraint on the effective action is given by
\beqa
S_{\rm eff}(\psi)-S_{\rm eff}(\psi')=\int d^{D-1}x \sqrt{-\bg}\nabla_a\Big[e^{-2\bphi}J^a(\psi)\Big],\label{TS}
\eeqa
where $J^a$ is an arbitrary covariant vector composed of the $(D-1)$-dimensional base space fields. The above constraint may be used to fix the coupling constants in the effective action ${\bf S}_{\rm eff}$ up to overall factors at each order of derivatives.

In the heterotic theory, the reduced action $S_{\rm eff}$ and the $\mathbb{Z}_2$-transformations in the reduction scheme \reef{reduc} are nonlinear in the YM scalar field $\alpha^{ij}$. In this paper, we speculate that if we keep track of only the zeroth and first order of the scalar field $\alpha^{ij}$ in both the reduced action $S_{\rm eff}$ and the $\mathbb{Z}_2$-transformations, and assume this field is constant, then the corresponding truncated T-duality constraint \reef{TS} still has enough constraints to fix the coupling constants in ${\bf S}_{\rm eff}$. Therefore, we divide the reduced action and the $\mathbb{Z}_2$-transformations $\psi'$ into two parts as
\beqa
S_{\rm eff}=S^L_{\rm eff}+S^{NL}_{\rm eff} \quad ; \quad \psi'=\psi^L+\psi^{NL},
\eeqa
where the upper index $L$ stands for the zeroth and linear order of the constant $\alpha^{ij}$, and the upper index $NL$ stands for the nonlinear orders of $\alpha^{ij}$. Then we truncate the nonlinear parts and propose that the following truncated constraint:
\beqa
S^L_{\rm eff}(\psi)-S^L_{\rm eff}(\psi^L)=\int d^{9}x \sqrt{-\bg}\nabla_a\Big[e^{-2\bphi}J^a(\psi)\Big],\label{TSL}
\eeqa
may fix all coupling constants in the parent action ${\bf S}_{\rm eff}$ up to overall factors at each derivative order. In the second term of the above action, only the zeroth and first order terms of $\alpha^{ij}$ should be retained.

The Buscher rules in \reef{bucher} only include the zeroth and first order terms of $\alpha^{ij}$. This part is necessary to investigate the T-duality of the reduced action at the zeroth and linear order of $\alpha^{ij}$. The nonlinear part of the Buscher rule, which we are not concerned with, is required to demonstrate the invariance of the full reduced action \reef{two3}, which incorporates terms up to fourth order in $\alpha^{ij}$, under T-duality \cite{Bergshoeff:1995cg}.
Assuming that the diffeomorphism transformations do not receive higher-derivative deformations, the Buscher rules must be generalized to include higher derivative terms corresponding to the effective action at each order of $\alpha'$. Therefore, the truncated generalized Buscher rules $\psi^L$ should have the following expansion:
\beqa
\psi^L=\psi_0^L+\sum_{n=1}^{\infty}\frac{\alpha'^n}{n!}\psi_n^L. \label{gBuch}
\eeqa
Here, $\psi_0^L$ represents the truncated Buscher rules in \reef{bucher}, and $\psi_n^L$ represents their deformations at order $\alpha'^n$. The deformed Buscher rules must form the $\mathbb{Z}_2$-group.

 Using an $\alpha'$-expansion for the $9$-dimensional effective action $S^L_{\rm eff}$ and for the vector $J^a$, and employing the following Taylor expansion for the T-duality transformation of the $9$-dimensional effective action at order $\alpha'^n$ around the truncated Buscher rule $\psi_0^L$:
\begin{equation}
S^{L(n)}(\psi^L)=\sum_{m=0}^{\infty}\alpha'^mS^{L(n,m)}(\psi_0^L),
\end{equation}
where $S^{L(n,0)}=S^{L(n)}$, one finds that the $\mathbb{Z}_2$-constraint in \reef{TSL} becomes
\beqa
\sum^\infty_{n=0}\alpha'^nS^{L(n)}(\psi)-\sum^\infty_{n=0,m=0}\alpha'^{n+m}S^{L(n,m)}(\psi_0^L)-\sum^\infty_{n=0}\alpha'^n\int d^{9}x\, \partial_a\Big[e^{-2\bphi}J^a_{(n)}(\psi)\Big]=0, \label{TSn}
\eeqa
where $J^a_{(n)}$ is an arbitrary covariant vector composed of the $9$-dimensional base space fields at order $\alpha'^{n+1/2}$. In the above equation, we have also used the observation made in \cite{Garousi:2019mca} that the $\mathbb{Z}_2$-constraint for curved base space produces exactly the same constraint for the coupling constants as for flat base space. Hence, for simplicity of the calculation, we have assumed the base space is flat. 
To find the appropriate constraints on the effective actions, one must set the terms at each order of $\alpha'$ to zero.

The T-duality constraint \reef{TSn} at order $\alpha'^0$ is
\beqa
S^{L(0)}(\psi)-S^{L(0)}(\psi^L_0)- \int d^{9}x\, \partial_a\Big[e^{-2\bphi}J_{(0)}^a(\psi)\Big]=0,\label{TS0}
\eeqa
where $S^{L(0)}(\psi)$ is determined by the terms in the first and second lines of \reef{two3}. We have already observed that the above constraint fixes the coupling constants in $\mathbf{S}^{(0)}$.

The constraint in \reef{TSn} at order $\alpha'$ becomes
\beqa
-S^{L(0,1)}(\psi_0^L)- \int d^{9}x\, \partial_a\Big[e^{-2\bphi}J^a_{(1)}(\psi)\Big]=S^{L(1)}(\psi_0^L)-S^{L(1)}(\psi).\label{TS3}
\eeqa
Using the reductions in \reef{reduc}, it is straightforward to find the circular reduction of the couplings in \reef{min} or \reef{max} to obtain the corresponding $S^{L(1)}(\psi)$. Then, using the truncated Buscher transformations in \reef{bucher}, one can calculate its corresponding $S^{L(1)}(\psi_0^L)$.

To determine the first term on the left-hand side of the above equation, we need higher-derivative corrections to the truncated Buscher rules \reef{bucher}, i.e.,
\beqa
&&\varphi^L= -\varphi+\alpha'\Delta\varphi^{(1)}(\psi)+\cdots
\,,\quad g^L_{a}= b_{a}+\alpha'e^{\varphi/2}\Delta g^{(1)}_a(\psi)+\cdots
\,,\quad\nn\\
&&b^L_{a}= g_{a}+\alpha'e^{-\varphi/2}\Delta b^{(1)}_a(\psi)+\cdots
\,,\quad \bar{g}_{ab}^L=\eta_{ab}+\alpha'\Delta \bar{g}^{(1)}_{ab}(\psi)+\cdots
\,,\quad\nn\\
&&\bar{H}_{abc}^L=\bar{H}_{abc}+\alpha'\Delta\bar{H}^{(1)}_{abc}(\psi)+\cdots
\,,\quad\bar{\phi}^L= \bar{\phi}+\alpha'\Delta\bar{\phi}^{(1)}(\psi)+\cdots\,,\nn\\
&&(\bar{A}_a^L)^{ij}=\bar{A}_a{}^{ij}+\alpha' \Delta\bar{A}_a^{(1)}{}^{ij}(\psi)+\cdots\,,\quad(\alpha^L)^{ij}=-\alpha^{ij}+\alpha' \Delta\alpha^{(1)}{}^{ij}(\psi)+\cdots.
\label{T22}
\eeqa
The corrections $\Delta\varphi^{(1)}$, $\Delta g^{(1)}_a$, $\Delta b^{(1)}_a$, $\Delta \bar{g}^{(1)}_{ab}$, $\Delta\bar{\phi}^{(1)}$, $\Delta\bar{A}_a^{(1)}{}^{ij}$, $\Delta\alpha^{(1)}{}^{ij}$ are independent and can be written as contractions of the base space fields at the two-derivative order. However, the correction $\Delta\bar{H}^{(1)}_{abc}$ is related to the corrections $\Delta b^{(1)}_a$, $\Delta \bar{g}^{(1)}_{ab}$, $\Delta\bar{A}_a^{(1)}{}^{ij}$, and an arbitrary two-form $\Delta B_{ab}^{(1)}$ that is also at the two-derivative order. This relation is a result of the Bianchi identity of $\bar{H}^L$, which is
\beqa
d\bar{H}^L+\frac{3}{2}dg^L\wedge db^L+\frac{3}{4}d\bar{A}^{Lij}\wedge d\bar{A}_L{}_{ij}=0.
\eeqa
The replacement \reef{T22} produces the following relation at order $\alpha'$:
\beqa
d\left(\Delta \bar{H}^{(1)}+3e^{\varphi/2}dg\wedge\Delta g^{(1)}+3e^{-\varphi/2}db\wedge \Delta b^{(1)}+3d\bar{A}^{ij}\wedge \Delta \bar{A}^{(1)}{}_{ij}\right)=0.
\eeqa
The expression inside the parentheses should be an exact three-form. Hence,
\beqa
\Delta \bar{H}^{(1)}_{abc}&=&3\partial_{[a}\Delta B^{(1)}_{bc]}-3e^{\varphi/2}V_{[ab}\Delta g^{(1)}_{c]}-3e^{-\varphi/2}W_{[ab}\Delta b^{(1)}_{c]}-3\bar{F}_{[ab}{}^{ij}\Delta \bar{A}^{(1)}_{c]ij},
\eeqa
where $3\partial_{[a}\Delta B^{(1)}_{bc]}$ is the  exact three-form.

Using the above relation, one finds the Taylor expansion of the T-duality transformation of the leading-order action around the truncated Buscher transformations to be:
\beqa
 S^{L(0,1)}(\psi_0^L)
 &\!\!\!\!\!=\!\!\!\!\!& -\frac{2}{\kappa'^2}\int d^{9}x e^{-2\bphi} \,  \Big[ \Big(\!\frac{1}{4}\prt^a\vp\prt^b\vp-2\prt^a\prt^b\bphi\!+\!\frac{1}{4}\bH^{acd}\bH^b{}_{cd} \!+\!\frac{1}{2}(e^\vp V^{ac}V^b{}_{c}\!+\!e^{-\vp} W^{ac}W^b{}_{c})\nn\\
 &&+\bF^{acij}\bF^b{}_{cij}+(e^{\vp/2}V^{ac}-e^{-\vp/2}W^{ac})\bF^b{}_{cij}\alpha^{ij}\Big)\Delta\bar{g}_{ab}^{(1)}+\Big(2\prt_c\prt^c\bphi-2\prt_c\bphi\prt^c\bphi -\frac{1}{24}\bH^2\nn\\&\!\!\!\!\!\!\!&-\frac{1}{8}\prt_c\vp\prt^c\vp-\frac{1}{8}(e^{\vp} V^2+e^{-\vp}W^2)+\frac{1}{2}\bF_{abij}\bF^{abij}+(e^{\vp/2}V^{ab}-e^{-\vp/2}W^{ab})\bF_{abij}\alpha^{ij}\Big)\nn\\&\!\!\!\!\!\!\!&\times(\eta^{ab}\Delta\bar{g}_{ab}^{(1)}-4\Delta\bphi^{(1)})-\Big(\frac{1}{2}\prt_a\prt^a\vp-\prt_a\bphi\prt^a\vp+\frac{1}{4}(e^{\vp/2}V^{ab}+e^{-\vp/2}W^{ab})\bF_{abij}\alpha^{ij}\nn\\&\!\!\!\!\!\!\!&-\frac{1}{4}(e^\vp V^2-e^{-\vp}W^2)\Big)\Delta\vp^{(1)}+\Big(2e^{-\vp/2}\prt_b\bphi W^{ab}- e^{-\vp/2}\prt_b W^{ab}+e^{-\vp/2}\prt_b\vp W^{ab}\nn\\&&+\frac{1}{2}e^{\vp/2}\bH^{abc}V_{bc}-2(\prt_b\bphi+\frac{1}{4}\prt_b\vp)\bF^{abij}\alpha_{ij}+\prt_b\bF^{abij}\alpha_{ij}\Big)\Delta g_a^{(1)} +\Big(2e^{\vp/2}\prt_b\bphi V^{ab}\nn\\&&- e^{\vp/2}\prt_b V^{ab}-e^{\vp/2}\prt_b\vp V^{ab}+\frac{1}{2}e^{-\vp/2}\bH^{abc}W_{bc}+2(\prt_b\bphi-\frac{1}{4}\prt_b\vp)\bF^{abij}\alpha_{ij}\nn\\&\!\!\!\!\!\!\!&-\prt_b\bF^{abij}\alpha_{ij}\Big)\Delta b_a^{(1)}+\Big(\frac{1}{2}\prt_a\bH^{abc}-\bH^{abc}\prt_a\bphi\Big)\Delta B_{bc}^{(1)}+\Big(\frac{1}{2}\bH^{abc}\bF_{bcij}-\prt_b\bF^{ab}{}_{ij}\nn\\&\!\!\!\!\!\!\!&+2\prt_b\bphi\bF^{ab}{}_{ij}+\alpha_{ij}[2(\prt_b\bphi-\frac{1}{4}\prt_b\vp)e^{\vp/2}V^{ab}-2(\prt_b\bphi+\frac{1}{4}\prt_b\vp)e^{-\vp/2}W^{ab}-e^{\vp/2}\prt_b V^{ab}\nn\\&\!\!\!\!\!\!\!&+e^{-\vp/2}\prt_bW^{ab}]\Big)\Delta \bA_a^{(1)ij}+\Big(\frac{1}{2}\bF_{abij}(e^{\vp/2}V^{ab}-e^{-\vp/2}W^{ab})\nn\\&\!\!\!\!\!\!\!&+\alpha_{ij}(-\frac{1}{2}V^{ab}W_{ab}+\frac{1}{2}W^2+\frac{1}{2}\bF^2+\frac{1}{4}\prt_a\vp\prt^a\vp)\Big)\Delta \alpha^{(1)ij}\Big] \,,\labell{d3S}
 \eeqa
where we have also removed some total derivative terms. In finding the above result in the last line, we have kept the terms in the leading-order action that have a second order of the scalar $\alpha^{ij}$. This is because the variation $\Delta \alpha^{(1)ij}$ may have terms at the zeroth order of $\alpha^{ij}$, which would produce a linear order of $\alpha^{ij}$ after replacing it in the last line above.

 Since the truncated T-duality transformations \reef{T22} must satisfy the $\mathbb{Z}_2$-group, the deformations at order $\alpha'$ should satisfy the following constraint:
\beqa
-\Delta\varphi^{(1)}(\psi)+\Delta\varphi^{(1)}(\psi_0^L) =0 &;&
\Delta b_a^{(1)}(\psi)+\Delta g_a^{(1)}(\psi_0^L) =0\,,\nn\\
\Delta g_a^{(1)}(\psi)+\Delta b_a^{(1)}(\psi_0^L)=0 &;&
\Delta \bar{g}_{ab}^{(1)}(\psi)+\Delta \bar{g}_{ab}^{(1)}(\psi_0^L)=0\,,\nn\\
\Delta\bar{\phi}^{(1)}(\psi)+\Delta\bar{\phi}^{(1)}(\psi_0^L) =0 &;&
\Delta B_{ab}^{(1)}(\psi)+\Delta B_{ab}^{(1)}(\psi_0^L)=0\,,\nn\\
\Delta \bar{A}_a^{(1)ij}(\psi)+\Delta \bar{A}_a^{(1)ij}(\psi_0^L)=0 &;&
-\Delta \alpha^{(1)ij}(\psi)+\Delta \alpha^{(1)ij}(\psi_0^L)=0\,.\label{Z22}
\eeqa
These corrections should be constructed from all contractions of $\partial \varphi$, $\partial \bar{\varphi}$, $e^{\varphi/2}V$, $e^{-\varphi/2}W$, $\bar{H}$, $\bar{F}_{ab}{}^{ij}$, and their derivatives at order $\alpha'$ with arbitrary coefficients. They should also include terms at zeroth and first order in the scalar $\alpha^{ij}$, while satisfying the above constraint.
 
Having discussed how to calculate $S^{L(0,1)}(\psi_0^L)$, let us now consider the total derivative term in the T-duality constraint \reef{TS3}. The vector $J^a_{(1)}$ should also be constructed from contractions of $\partial \varphi$, $\partial \bar{\varphi}$, $e^{\varphi/2}V$, $e^{-\varphi/2}W$, $\bar{H}$, $\bar{F}_{ab}{}^{ij}$ at the three-derivative order with arbitrary coefficients. Furthermore, these constructions should also  include terms at zeroth and first order in the scalar $\alpha^{ij}$.

The parameters in the total derivative term and in the corrections to the Buscher rules appear on the left-hand side of the equation \reef{TS3}, while the coupling constants $c_1, c_2, \ldots$ appear on the right-hand side of this equation. To solve this equation, we need to impose the Bianchi identities corresponding to the field strengths $\bH, \bF, V, W$. We impose these identities by expressing the field strengths in terms of potentials. As a result, the equation can be written in terms of independent but non-gauge-invariant couplings. The coefficients of the independent terms must be zero, leading to a system of linear algebraic equations among these parameters.

By solving the algebraic equations, we obtain two sets of solutions for these parameters. One set of solutions represents the relationships only among the parameters in the deformations of the Buscher rules and total derivative terms. These solutions satisfy the following homogeneous equation:
\begin{equation}
-S^{L(0,1)}(\psi_0^L)-\int d^{9}x\, \partial_a\left[e^{-2\bar{\phi}}J^a_{(1)}(\psi)\right]=0. \label{TSB3h}
\end{equation}
However, we are not interested in the solutions of the above equation.

The other solution, which is a particular solution of the non-homogeneous equation \reef{TS3} and the one of interest, expresses these parameters in terms of the coupling constants $c_1, c_2, \ldots$. This solution also determines the relationships between the coupling constants. We will consider the particular solution for two cases: when the couplings are in the minimal basis and when the couplings are in the maximal basis. In the next subsection, we will focus on the case of the minimal basis.

\subsection{Couplings in the Minimal Scheme}

If we consider the four-derivative couplings in the minimal basis \reef{min} without the inclusion of the Chern-Simons couplings $H\Omega$, the particular solution of the non-homogeneous equation \reef{TS3} yields $c_1=c_2=\cdots=c_{24}=0$. This result is expected since there is no string theory in which $H$ is given by \reef{FH}. 
However, in heterotic theory, the Green-Schwarz mechanism \cite{Green:1984sg} introduces the YM Chern-Simons term in \reef{FH} as well as the Lorentz Chern-Simons terms $-(3\alpha'/2) \Omega_{\mu\nu\rho}$. These terms give rise to an additional four-derivative coupling, along with the 24 couplings in the minimal basis. Interestingly, when we include this term, the T-duality constraint \reef{TS3} fixes all the parameters in terms of the coefficient of the Chern-Simons coupling $H\Omega$, which is already fixed at the two-derivative order.

We find the following result for the effective action:
\beqa
{\bf S}^{(1)}&=&-\frac{2\alpha' }{8\kappa^2}\int d^{10}x \sqrt{-G}e^{-2\Phi}\Big[\frac{1}{4} F_{\alpha }{}^{\gamma kl} F^{\alpha \beta ij} F_{\beta }{}^{\delta }{}_{kl} F_{\gamma \delta ij} -  \frac{1}{2} F_{\alpha }{}^{\gamma }{}_{ij} F^{\alpha \beta ij} F_{\beta }{}^{\delta kl} F_{\gamma \delta kl}\nn\\&& -  \frac{1}{8} F_{\alpha \beta }{}^{kl} F^{\alpha \beta ij} F_{\gamma \delta kl} F^{\gamma \delta }{}_{ij} + \frac{1}{4} F^{\alpha \beta ij} F^{\gamma \delta }{}_{ij} H_{\alpha \gamma }{}^{\epsilon } H_{\beta \delta \epsilon } -  \frac{1}{8} F^{\alpha \beta ij} F^{\gamma \delta }{}_{ij} H_{\alpha \beta }{}^{\epsilon } H_{\gamma \delta \epsilon }\nn\\&& -  \frac{1}{2} F_{\alpha }{}^{\gamma }{}_{ij} F^{\alpha \beta ij} H_{\beta }{}^{\delta \epsilon } H_{\gamma \delta \epsilon } + \frac{1}{24} H_{\alpha }{}^{\delta \epsilon } H^{\alpha \beta \gamma } H_{\beta \delta }{}^{\varepsilon } H_{\gamma \epsilon \varepsilon } -  \frac{1}{8} H_{\alpha \beta }{}^{\delta } H^{\alpha \beta \gamma } H_{\gamma }{}^{\epsilon \varepsilon } H_{\delta \epsilon \varepsilon }  \nn\\&&+ R_{\alpha \beta \gamma \delta } R^{\alpha \beta \gamma \delta }-  \frac{1}{2} H_{\alpha }{}^{\delta \epsilon } H^{\alpha \beta \gamma } R_{\beta \gamma \delta \epsilon }+2H^{\alpha\beta\gamma}\Omega_{\alpha\beta\gamma}\Big]\,.\labell{fourmin}
\eeqa
Interestingly, all unambiguous single-trace four YM couplings become zero, which is consistent with the results obtained from the S-matrix method \cite{Gross:1986mw}. When the YM gauge field is zero, the action mentioned above reduces to  the Metsaev-Tseytlin action, which is consistent with the S-matrix elements in the heterotic theory \cite{Metsaev:1987zx}.

The T-duality constraint \reef{TS3} also yields the following corrections to the truncated  Buscher rules \reef{bucher}:
\beqa
-8 \Delta\bphi^{(1)}&=& - \frac{1}{2} e^{\vp} V_{ab} V^{ab} +\frac{1}{2}{ e^{-\vp}}W_{ab} W^{ab}\,,\nn\\
-8\Delta G_{ab}^{(1)}&=& -2 e^{\vp} V_{a}{}^{c} V_{bc} +2{e^{-\vp}} W_{a}{}^{c} W_{bc}\,,\nn\\
-8\Delta\alpha_{ij}^{(1)}&=&- \frac{1}{2} e^{\vp} V_{cd} V^{cd} \alpha_{ij} + \frac{1}{2}{e^{-\vp}}W_{cd} W^{cd} \alpha_{ij}\,,\nn\\
-8\Delta  \bA_a^{(1)}{}_{ij}&=& - e^{ \vp/2} \bH_{ade} V^{de} \alpha_{ij} -{e^{\vp/2}} \bH_{ade} W^{de} \alpha_{ij}\,,\nn\\
-8\Delta B_{ab}^{(1)}&=& 4 V_{[b}{}^{c} W_{a]c} +2 e^{\vp/2} \bF_{[b}{}^{cij} V_{a]}{}_{c} \alpha_{ij}  +2{e^{- \vp/2}}\bF_{[b}{}^{cij} W_{a]}{}_{c} \alpha_{ij}\,,\nn\\
-8\Delta\vp^{(1)}&=& -2 e^{\vp} V_{ab} V^{ab}  - 2{e^{-\vp}} W_{ab} W^{ab} - 2 \prt_{a}\vp \prt^{a}\vp+ 2 V^{ab} W_{ab}\,,\nn\\
-8\Delta g_a^{(1)}&=&- e^{ \vp/2} \bH_{abc} V^{bc}-  2{e^{-\vp/2}}\prt_{b}W_{a}{}^{b} + 4 {e^{- \vp/2}}W_{ab}\prt^{b}\bphi +\frac{1}{2}e^{-\vp/2}\bH_{abc}W^{bc}-e^{\vp/2}V_{ab}\prt^b\vp\nn\\&&+\alpha_{ij}\Big(\frac{1}{2}\bH_{abc}\bF^{bcij}-2\prt_a\bF_a{}^{bij}+4\bF_{ab}{}^{ij}\prt^b\bphi+\bF_{ab}{}^{ij}\prt^b\vp\Big)\,,\labell{minrule}
\eeqa
and $\Delta b_a^{(1)}(\psi)=-\Delta g_a^{(1)}(\psi_0^L)$. When the YM scalar $\alpha^{ij}$ is zero, the above correction reduces to the corrections that have been found in \cite{Garousi:2019wgz} for the couplings in the  Metsaev-Tseytlin action.
 
The observation that the single-trace terms of the four YM field strength are zero is consistent with the S-matrix results. However, the two-trace terms in \reef{fourmin} do not precisely match those obtained from S-matrix calculations \cite{Gross:1986mw}. This discrepancy is not an inconsistency because the two-trace terms represent ambiguous  couplings that can be modified through field redefinitions. Therefore, the explicit form of the four YM field strength that is produced by the S-matrix method should appear in other schemes. To confirm the consistency of the two-trace terms with the S-matrix, we need to derive the effective action in alternative schemes. In the next subsection, we will determine the couplings in the maximal scheme.

\subsection{Couplings in the Maximal Scheme}

If we consider the four-derivative couplings in the maximal basis \reef{max} without the inclusion of the Chern-Simons couplings $H\Omega$, the particular solution of the non-homogeneous equation \reef{TS3} yields 24 relations among the 42 coupling constants. We then examine the resulting basis, which contains 18 parameters under the field redefinition. Among these parameters, $c_{23}$ in the basis \reef{max}, which represents the coefficient of the Riemann squared, remains unchanged under field redefinitions. The remaining 17 parameters can be eliminated through suitable field redefinitions.
If there were a string theory in which the Lorentz Chern-Simons term in the $B$-field strength was absent, the unambiguous parameter $c_{23}$ would be non-zero and represent the coefficient of four-derivative couplings. However, since no such theory exists, we must impose the condition that the unambiguous parameter must be zero. This condition introduces an additional constraint on top of the 24 relations obtained from the T-duality constraint \reef{TS3}.

However, in the heterotic theory, there exist four-derivative Chern-Simons couplings $H\Omega$ along with the 42 couplings in the maximal basis. In this case, the particular solution of the non-homogeneous equation \reef{TS3} yields the following 24 relations between the 42 coupling constants and the coefficient of the couplings $H\Omega$:
\beqa
&\!\!\!\!\!\!&c_{12} = 1/24 + c_{1}/12 + c_{10}/6, 
c_{18} = -1 - 8 c_{1} - 8 c_{10} - 8 c_{11} + 16 c_{14} - 2 c_{16}, 
c_{2} = 0,\nn\\&\!\!\!\!\!\!& c_{22} = -1/4 - 2 c_{10}, c_{23} = 1 + 2 c_{1} + 4 c_{10}, 
c_{24} = -3/4 - 2 c_{10} + 2 c_{11} - 8 c_{14} + c_{16}/2 - c_{17}, \nn\\&\!\!\!\!\!\!&
c_{25} = -1/2 + c_{1} - 2 c_{10}, c_{3} = 0, 
c_{31} = 32 c_{19} - 48 c_{20} + 8 c_{21} - 16 c_{26} + 24 c_{27} - 
  2 c_{28} - 8 c_{29} + 12 c_{30}, \nn\\&\!\!\!\!\!\!&
c_{32} = 4 + 32 c_{1} + 32 c_{10} + 32 c_{11} + 96 c_{13} - 64 c_{14} - 
  432 c_{15} + 8 c_{16} + 40 c_{19} - 108 c_{20} - 15 c_{21} - 8 c_{26}\nn\\&\!\!\!\!\!\!& + 
  24 c_{27} + 4 c_{28}, c_{34} = c_{33}/2, 
c_{35} = -4 - 32 c_{1} - 32 c_{10} - 32 c_{11} + 64 c_{14} - 8 c_{16} - 
  2 c_{33}, \nn\\&\!\!\!\!\!\!&c_{36} = -16 - 128 c_{1} - 128 c_{10} - 128 c_{11} - 
  384 c_{13} + 256 c_{14} + 1728 c_{15} - 32 c_{16} - 32 c_{19} + 
  240 c_{20}\nn\\&\!\!\!\!\!\!& + 76 c_{21} - 32 c_{26} - 16 c_{28} - 32 c_{29} + 48 c_{30}, 
c_{37} = 24 + 192 c_{1} + 192 c_{10} + 192 c_{11} + 768 c_{13} - 
  384 c_{14}\nn\\&\!\!\!\!\!\!& - 3456 c_{15} + 48 c_{16} - 288 c_{20} - 120 c_{21} + 
  96 c_{26} - 96 c_{27} + 24 c_{28} + 80 c_{29} - 144 c_{30} - 4 c_{33}, \nn\\&\!\!\!\!\!\!&
c_{38} = -2 - 16 c_{1} - 16 c_{10} - 16 c_{11} + 96 c_{13} + 32 c_{14} - 
  288 c_{15} - 2 c_{16} + 40 c_{19} - 48 c_{20} + 10 c_{21} - 8 c_{26}\nn\\&\!\!\!\!\!\!& + 
  12 c_{27} - c_{28}, c_{4} = 2 c_{11} - 4 c_{14}, c_{40} = 1/2 - 2 c_{1}, 
c_{41} = 1 + 8 c_{1} + 8 c_{10} + 8 c_{11} - 48 c_{13} - 16 c_{14}\nn\\&\!\!\!\!\!\!& + 
  144 c_{15} + 2 c_{16} - 2 c_{17} - 20 c_{19} + 24 c_{20} - 5 c_{21} + 
  4 c_{26} - 6 c_{27} + c_{28}/2 + c_{39}, 
c_{42} = 1/4 - c_{1} \nn\\&\!\!\!\!\!\!&- 2 c_{11} + 8 c_{14} - c_{16}/2 + c_{17}, 
c_{5} = -(c_{1}/4) + c_{10}/2, c_{6} = 0, c_{7} = 0, 
c_{8} = (3 c_{13})/2 - (9 c_{15})/4\nn\\&\!\!\!\!\!\!& - c_{19}/8 + (3 c_{20})/16 - c_{21}/64,
 c_{9} = 1/4 + c_{1}/2 + c_{10}\,.\labell{cons1}
\eeqa
Using the fact that the coefficient of Riemann squared, $c_{23}$, must be zero in the absence of the couplings $H\Omega$, we obtain one additional relation:
\begin{equation}
c_{10}=-\frac{c_1}{2}\,.\label{cons2}
\end{equation}
By substituting the relations \reef{cons1} and \reef{cons2} into the maximal basis \reef{max}, we determine the couplings in the maximal scheme, which involve 17 arbitrary parameters. Choosing specific values for these parameters allows us to obtain the effective action in the corresponding scheme. For instance, when $c_{1}=1/4$, $c_{11}=-1/2$, $c_{14}=-1/8$, and all other 14 parameters are set to zero, we recover the action \reef{fourmin}. 

In order to compare the four YM couplings generated by T-duality with the corresponding couplings obtained from the S-matrix method, we consider a scheme in which the graviton, dilaton, and $B$-field propagators derived from the leading-order action \reef{two} receive no corrections at the four-derivative order. One particular scheme satisfying this condition is the Meissner action. With the following relations for the 17 parameters:
\beqa
&\!\!\!\!\!&c_1=1/4,c_{14}=-1/8,c_{21}=1,c_{15}=1/144,c_{17}=1,c_{20}=-1/6, c_{27}=-2/3,c_{30}=2/3,c_{39}=2,\nn\\&\!\!\!\!\!&
c_{16}=c_{19}=c_{11}=c_{28}=c_{29}=c_{26}=c_{33}=c_{13}=0\,,
\eeqa 
we obtain the following effective action:
\beqa
{\bf S}^{(1)}&=&-\frac{2\alpha' }{8\kappa^2}\int d^{10}x \sqrt{-G}e^{-2\Phi}\Big[\frac{1}{4} F_{\alpha }{}^{\gamma kl} F^{\alpha \beta ij} 
F_{\beta }{}^{\delta }{}_{kl} F_{\gamma \delta ij} + 
\frac{1}{2} F_{\alpha }{}^{\gamma }{}_{ij} F^{\alpha \beta 
ij} F_{\beta }{}^{\delta kl} F_{\gamma \delta kl}\nn\\&& -  
\frac{1}{8} F_{\alpha \beta }{}^{kl} F^{\alpha \beta ij} F_{
\gamma \delta kl} F^{\gamma \delta }{}_{ij} -  \frac{1}{16} 
F_{\alpha \beta ij} F^{\alpha \beta ij} F_{\gamma \delta kl} 
F^{\gamma \delta kl} + \frac{1}{4} F^{\alpha \beta ij} 
F^{\gamma \delta }{}_{ij} H_{\alpha \gamma }{}^{\epsilon } 
H_{\beta \delta \epsilon } \nn\\&&-  \frac{1}{8} F^{\alpha \beta ij} 
F^{\gamma \delta }{}_{ij} H_{\alpha \beta }{}^{\epsilon } 
H_{\gamma \delta \epsilon } + \frac{1}{24} H_{\alpha 
}{}^{\delta \epsilon } H^{\alpha \beta \gamma } H_{\beta 
\delta }{}^{\varepsilon } H_{\gamma \epsilon \varepsilon } -  
\frac{1}{8} H_{\alpha \beta }{}^{\delta } H^{\alpha \beta 
\gamma } H_{\gamma }{}^{\epsilon \varepsilon } H_{\delta 
\epsilon \varepsilon } \nn\\&&+ \frac{1}{144} H_{\alpha \beta \gamma 
} H^{\alpha \beta \gamma } H_{\delta \epsilon \varepsilon } H^{
\delta \epsilon \varepsilon } + H_{\alpha }{}^{\gamma \delta } 
H_{\beta \gamma \delta } R^{\alpha \beta } - 4 
R_{\alpha \beta } R^{\alpha \beta } -  
\frac{1}{6} H_{\alpha \beta \gamma } H^{\alpha \beta \gamma } 
R + R^2 \nn\\&&+ R_{\alpha \beta \gamma 
\delta } R^{\alpha \beta \gamma \delta } -  
\frac{1}{2} H_{\alpha }{}^{\delta \epsilon } H^{\alpha \beta 
\gamma } R_{\beta \gamma \delta \epsilon } -  
\frac{2}{3} H_{\beta \gamma \delta } H^{\beta \gamma \delta } 
\nabla_{\alpha }\nabla^{\alpha }\Phi + \frac{2}{3} H_{\beta 
\gamma \delta } H^{\beta \gamma \delta } \nabla_{\alpha }\Phi 
\nabla^{\alpha }\Phi \nn\\&&+ 8 R \nabla_{\alpha }\Phi 
\nabla^{\alpha }\Phi - 16 R_{\alpha \beta } 
\nabla^{\alpha }\Phi \nabla^{\beta }\Phi + 16 \nabla_{\alpha 
}\Phi \nabla^{\alpha }\Phi \nabla_{\beta }\Phi \nabla^{\beta 
}\Phi - 32 \nabla^{\alpha }\Phi \nabla_{\beta }\nabla_{\alpha 
}\Phi \nabla^{\beta }\Phi \nn\\&&+ 2 H_{\alpha }{}^{\gamma \delta } 
H_{\beta \gamma \delta } \nabla^{\beta }\nabla^{\alpha }\Phi +2H^{\alpha\beta\gamma}\Omega_{\alpha\beta\gamma}\Big]\,.\labell{fourmax}
\eeqa
When the YM gauge field is zero, the action mentioned above reduces to the Meissner action \cite{Meissner:1996sa}, up to the  total derivative term $16\nabla_\beta(e^{-2\Phi}\nabla_\alpha\Phi\nabla^\alpha\Phi\nabla^\beta\Phi)$. The corresponding corrections to the truncated  Buscher rules \reef{bucher} are:
\beqa
-8\Delta\bphi^{(1)}&=& - \frac{1}{2} e^{\vp/2} \bF_{abij} V^{ab} \alpha^{ij} -  \frac{1}{2}{ e^{-\vp/2}}\bF_{abij} W^{ab} \alpha^{ij}\,,\labell{maxrule} \\
-8\Delta G_{ab}^{(1)}&=& -4 e^{\vp/2} \bF_{\{b}{}^{cij} V_{a\}}{}_{c} \alpha_{ij} - 4{e^{-\vp/2}}
\bF_{\{b}{}^{cij} W_{a\}}{}_{c} \alpha_{ij} \,,\nn\\
-8\Delta\alpha_{ij}^{(1)}&=&-  e^{\vp} V_{cd} V^{cd} \alpha_{ij} +{e^{-\vp}}W_{cd} W^{cd} \alpha_{ij}+ 2 \alpha_{ij} \prt_{c}\prt^{c}\vp - 4 \alpha _{ij}\prt_{c}\vp\prt^{c}\bphi\,,\nn\\
-8\Delta  \bA_a^{(1)}{}_{ij}&=& 0\,,\nn\\
-8\Delta B_{ab}^{(1)}&=& 4 V_{[b}{}^{c} W_{a]c} +2 e^{\vp/2} \bF_{[b}{}^{cij} V_{a]}{}_{c} \alpha_{ij}  +2{e^{- \vp/2}}\bF_{[b}{}^{cij} W_{a]}{}_{c} \alpha_{ij}\,,\nn\\
-8\Delta\vp^{(1)}&=&- e^{\vp} V_{ab} V^{ab}  - {e^{-\vp}} W_{ab} W^{ab} - 2 \prt_{a}\vp \prt^{a}\vp+ 2 V^{ab} W_{ab}\,,\nn\\
-8\Delta g_a^{(1)}&=&- e^{ \vp/2} \bH_{abc} V^{bc}-2 {e^{- \vp/2}}W_{ab}\prt^{b}\vp +\frac{1}{2}e^{-\vp/2}\bH_{abc}W^{bc}-e^{\vp/2}V_{ab}\prt^b\vp-\frac{1}{2}\bH_{abc}\bF^{bcij}\alpha_{ij}\,,\nn
\eeqa
and $\Delta b_a^{(1)}(\psi)=-\Delta g_a^{(1)}(\psi_0^L)$. When the YM scalar $\alpha^{ij}$ is set to zero, the aforementioned correction simplifies to the corrections previously discovered in \cite{Kaloper:1997ux, Garousi:2019wgz} for the couplings in the Meissner action. It is worth noting that while the base space metric and dilaton remain invariant for NS-NS couplings, the presence of the YM field breaks this invariance. Furthermore, in the absence of the YM field, the transformations only involve the first derivative of the base space fields, whereas the YM field introduces second derivatives of the base space fields in addition to the first derivative.

The four YM coupling in \reef{fourmax} can be expressed as follows:
\beqa
&&\frac{1}{4}\Tr(F_\alpha{}^\gamma F_\beta{}^{\delta})\Tr(F^{\alpha\beta}F_{\gamma\delta})+\frac{1}{2}\Tr(F_a{}^{\gamma}F^{\alpha\beta})\Tr(F_\beta{}^{\gamma}F_{\gamma\delta})\nn\\&&-\frac{1}{8}\Tr(F_{\alpha\beta}F_{\gamma\delta})\Tr(F^{\alpha\beta}F^{\gamma\delta})-\frac{1}{16}\Tr(F_{\alpha\beta}F^{\alpha\beta})\Tr(F_{\gamma\delta}F^{\gamma\delta})\nn\\&&=\frac{1}{32}t^{\alpha\beta\gamma\delta\mu\nu\rho\lambda}\Tr(F_{\alpha\beta}F_{\gamma\delta})\Tr(F_{\mu\nu}F_{\rho\lambda})\,.\labell{4YM}
\eeqa
 The tensor $t_8$, as defined in \cite{Schwarz:1982jn}, is such that the contraction of $t_8$ with four arbitrary antisymmetric tensors $M^1, \cdots, M^4$ is given by:
\beqa
t^{\alpha\beta\gamma\delta\mu\nu\rho\sigma}M^1_{\alpha\beta}M^2_{\gamma\delta}M^3_{\mu\nu}M^4_{\rho\sigma}&\!\!\!\!\!=\!\!\!\!\!&8(\tr M^1M^2M^3M^4+\tr M^1M^3M^2M^4+\tr M^1M^3M^4M^2)\labell{t8}\\
&&-2(\tr M^1M^2\tr M^3M^4+\tr M^1M^3\tr M^2M^4+\tr M^1M^4\tr M^2M^3)\nn
\eeqa
The  couplings in \reef{4YM} are precisely the four YM couplings that have been discovered in \cite{Gross:1986mw} through the examination of the low-energy expansion of the sphere-level S-matrix element involving four YM vertex operators in heterotic string theory. Furthermore, the couplings between the YM field strength and the $H$-field in \reef{fourmax} can be expressed as follows:
\beqa
\Big( \frac{1}{4}H_{\alpha\gamma}{}^{\epsilon}H_{\beta\delta\epsilon}-\frac{1}{8}H_{\alpha\beta}{}^{\epsilon}H_{\gamma\delta\epsilon}\Big)\Tr(F^{\alpha\beta}F^{\gamma\delta})&=&-\frac{3}{8}H_{[\alpha \beta}{}^{\epsilon}H_{\gamma]\delta\epsilon}\Tr(F^{\alpha\beta}F^{\gamma\delta})\,,
 \eeqa
which is also  the coupling found through the S-matrix method in \cite{Gross:1986mw}.
It is important to note that the couplings presented in \cite{Gross:1986mw} are in the Einstein frame, whereas the couplings in equation \reef{fourmax} are in the string frame. Furthermore, the coupling schemes in \cite{Gross:1986mw} and equation \reef{fourmax} differ from each other. However, in both cases, the leading-order propagators do not receive four-derivative corrections.

If one would like to compare the remaining couplings in \cite{Gross:1986mw} with those in equation \reef{fourmax}, it would be necessary to first convert from the Einstein frame to the string frame and then use field redefinition to precisely identify all terms in \cite{Gross:1986mw} with those in equation \reef{fourmax}. However, since the NS-NS couplings in the action \reef{fourmax} are the same as the couplings in the Metsaev-Tseytlin action, up to field redefinition, and the latter couplings are consistent with the S-matrix elements, we are not interested in delving into this comparison in further detail.

\section{Discussion}

In this paper, we have explored the imposition of the $O(1,1,\mathbb{Z})$ symmetry on the circular reduction of heterotic string theory in order to derive its effective action at the four-derivative level, focusing on NS-NS and YM couplings. Due to the nonlinear nature of T-duality transformations and the circular reduction of the effective action with respect to the scalar component of the YM field, we have introduced a truncation method. This method retains only the zeroth and first order terms of the scalar, discarding all higher order contributions, in both the T-duality transformations and the reduced action.

We have demonstrated that this truncation procedure can be employed within the T-duality $\mathbb{Z}_2$-constraint to determine the coupling constants in both the minimal basis and the maximal basis at the four-derivative order. In the minimal scheme, this leads to the inclusion of the YM coupling in the Metsaev-Tseytlin action, while in the maximal scheme, it allows for the incorporation of the YM couplings into the Meissner action.

We have obtained the truncated Buscher rules at two-derivative order, given by equations \reef{minrule} and \reef{maxrule}, which correspond to the Metsaev-Tseytlin action \reef{fourmin} and the Meissner action \reef{fourmax}, respectively. The circular reduction of these actions, considering only the zeroth and first order terms of the YM scalar, remains invariant under these truncated Buscher rules.
For the circular reduction of these actions at all orders of the YM scalar field, it is expected that they should be invariant under the full Buscher rules at two derivatives. These full Buscher rules encompass all higher orders of the YM scalar field, as well as derivatives of this scalar field. Discovering these transformations would be of great interest.

We have observed that the four YM field strength couplings, which are single-trace, are unambiguous because they cannot be generated through field redefinitions, Bianchi identities, or total derivative terms. However, the T-duality constraint requires their coefficients to be zero. To illustrate this more clearly, let's consider the circular reduction of the term in equation \reef{min} with coefficient $c_2$. By applying the reduction described in equation \reef{reduc}, we find that the reduced form of this term is
$
c_2 \, \bF_a{}^c{}_i{}^k \, \bF^{abij} \, \bF_b{}^d{}_k{}^l \, \bF_{cdjl} + 4c_2 \, e^{\vp/2} \, \bF_a{}^c{}_i{}^k \, \bF_b{}^d{}_k{}^l \, \bF_{cdjl} \, V^{ab} \, \alpha^{ij}\,.
$
The first term is invariant under the truncated Buscher rule stated in equation \reef{bucher}, while the second term transforms as 
$
- 4c_2 \, e^{-\vp/2} \, \bF_a{}^c{}_i{}^k \, \bF_b{}^d{}_k{}^l \, \bF_{cdjl} \, W^{ab} \, \alpha^{ij}\,.
$
No other term in equation \reef{min} produces such a structure. Furthermore, since it is a single-trace term, it cannot be generated by corrections to the Buscher rules, total derivative terms, or Bianchi identities in the base space. Therefore, T-duality fixes the coefficient $c_2$ to be zero. A similar situation occurs for the terms in equation \reef{min} with coefficients $c_3$, $c_6$, and $c_7$. We anticipate that a similar scenario would arise for higher-derivative couplings. Thus, we speculate that the classical effective action of the heterotic theory, at $n$ orders of derivatives and for $n>2$, should not contain single-trace couplings of $n$ YM field strengths. This is in contrast to the classical effective action of $D$-branes, which only features single-trace couplings of YM field strengths. This distinction arises from the fact that the vertex operator in the disk-level S-matrix element appears on the boundary of the disk, and there is a Chan-Paton factor for any S-matrix element. As a result, the classical effective action of type I string theory should only contain single-trace terms. This is not inconsistent with the S-duality between the heterotic theory and type I theory \cite{Witten:1995ex} because the higher-derivative classical effective action of one theory at weak coupling should be transformed into the higher-derivative effective action of the other theory at strong coupling, which is not a classical effective action.

The four YM couplings generated by T-duality in the Meissner scheme can be expressed in terms of the $t_8$ tensor, as shown in equation \reef{4YM}. On the other hand, T-duality yields the following eight-derivative coupling in the heterotic theory \cite{Razaghian:2018svg}:
\begin{equation}
\frac{\alpha'^2}{128}t^{\alpha\beta\gamma\delta\mu\nu\rho\lambda}\Tr(R_{\alpha\beta}R_{\gamma\delta})\Tr(R_{\mu\nu}R_{\rho\lambda}).
\end{equation}
Here, the trace is over the last two indices of the Riemann curvature. These two couplings can be combined as follows:
\begin{equation}
\frac{1}{32}t^{\alpha\beta\gamma\delta\mu\nu\rho\lambda}\Big[\Tr(F_{\alpha\beta}F_{\gamma\delta})-\frac{\alpha'}{2}\Tr(R_{\alpha\beta}R_{\gamma\delta})\Big]\Big[\Tr(F_{\mu\nu}F_{\rho\lambda})-\frac{\alpha'}{2}\Tr(R_{\mu\nu}R_{\rho\lambda})\Big],\label{FFRR}
\end{equation}
This expression has been determined in \cite{Gross:1986mw} through the study of the four-point S-matrix element.

The four-derivative couplings presented in equations \reef{fourmin} and \reef{fourmax} are consistent with the couplings obtained from the sphere-level four-point function in heterotic string theory. Calculating the six-point function in heterotic string theory is a highly intricate task, and extracting the corresponding low-energy six-derivative coupling is even more challenging. However, extending the approach used in this paper to the six-derivative order is relatively less complex compared to the S-matrix method. Therefore, it would be intriguing to apply the T-duality approach to uncover the six-derivative YM couplings in the heterotic theory, which should include the six-derivative couplings in \reef{FFRR}. It is worth noting that the NS-NS coupling at this order has already been derived using the T-duality method in a study by \cite{Garousi:2023kxw}.

 

\end{document}